\documentclass[journal]{IEEEtran}
\usepackage[T1]{fontenc}
\usepackage[utf8]{inputenc}
\usepackage{amssymb}
\usepackage[pdftex]{graphicx}
\usepackage{subcaption} 
\usepackage{amsmath}
\usepackage{amsthm}
\usepackage[cmintegrals]{newtxmath}
\usepackage{bm} 
\usepackage{array}
\usepackage[noadjust]{cite}
\usepackage{glossaries}
\usepackage{algorithm}
\usepackage{algpseudocode}
\usepackage{xcolor}
\usepackage{tikz}
\usepackage{pgfplots}
\usepackage{multirow}
\usepackage{makecell}
\usepackage{fixltx2e}

\newtheorem{thm}{Theorem}
\newtheorem{lmm}{Lemma}
\pgfdeclarelayer{background}
\pgfdeclarelayer{foreground}
\pgfsetlayers{background,main,foreground}
\usepgfplotslibrary{fillbetween} 
\usepgfplotslibrary{external} 
\newcommand{\rev}[1]{\textcolor{black}{{#1}}}

\makeatletter
\let\MYcaption\@makecaption
\makeatother

\usepackage[font=footnotesize]{subcaption}

\makeatletter
\let\@makecaption\MYcaption
\makeatother
\algnewcommand{\LineComment}[1]{\(\triangleright\) #1}

\newacronym{AWGN}{AWGN}{additive white Gaussian noise}
\newacronym{BER}{BER}{bit error rate}
\newacronym{BLER}{BLER}{block error rate}
\newacronym{BP}{BP}{backpropagation}
\newacronym{CE}{CE}{cross-entropy}
\newacronym{CFO}{CFO}{carrier frequency offset}
\newacronym{CP}{CP}{cyclic prefix}
\newacronym{CSI}{CSI}{channel state information}
\newacronym{DL}{DL}{deep learning}
\newacronym{DFT}{DFT}{discrete Fourier transform}
\newacronym{FFT}{FFT}{fast Fourier transform}
\newacronym{IFFT}{IFFT}{inverse fast Fourier transform}
\newacronym{GAN}{GAN}{generative adversarial network}
\newacronym{GPU}{GPU}{Graphics Processing Unit}
\newacronym{iid}{i.i.d.\@}{independent and identically distributed}
\newacronym{ISI}{ISI}{inter-symbol interference}
\newacronym{LOS}{LOS}{line-of-sight}
\newacronym{MDP}{MDP}{Markov decision process}
\newacronym{MIMO}{MIMO}{multiple-input multiple-output}
\newacronym{ML}{ML}{machine learning}
\newacronym{NN}{NN}{neural network}
\newacronym{OFDM}{OFDM}{orthogonal frequency-division multiplexing}
\newacronym{pdf}{pdf}{probability density function}
\newacronym{pmf}{pmf}{probability mass function}
\newacronym{RBF}{RBF}{Rayleigh block-fading}
\newacronym{ReLU}{ReLU}{rectified linear unit}
\newacronym{RL}{RL}{reinforcement learning}
\newacronym{SDR}{SDR}{software defined radio}
\newacronym{SER}{SER}{symbol error rate}
\newacronym{SNR}{SNR}{signal-to-noise ratio}
\newacronym{SINR}{SINR}{signal-to-interference-plus-noise ratio}
\newacronym{SGD}{SGD}{stochastic gradient descent}
\newacronym{wrt}{w.r.t.\@}{with respect to}
\newacronym{QPSK}{QPSK}{quaternary phase-shift keying}
\newacronym{SPSA}{SPSA}{simultaneous perturbation stochastic approximation}
\newacronym{SWIPT}{SWIPT}{simultaneous wireless information and power transfer}
\newacronym{MSE}{MSE}{mean squared error}
\newacronym{PSNR}{PSNR}{peak signal-to-noise ratio}
\newacronym{luib}{LUIB}{locally uniformly integrably bounded}

\renewcommand{\vec}[1]{\mathbf{#1}}
\newcommand{\vecs}[1]{\boldsymbol{#1}}

\newcommand{\gv}{\vec{g}}
\newcommand{\hv}{\vec{h}}

\newcommand{\lv}{\vec{l}}
\newcommand{\mv}{\vec{m}}
\newcommand{\nv}{\vec{n}}

\newcommand{\pv}{\vec{p}}

\newcommand{\wv}{\vec{w}}
\newcommand{\xv}{\vec{x}}
\newcommand{\yv}{\vec{y}}
\newcommand{\zv}{\vec{z}}
\newcommand{\zerov}{\vec{0}}

\newcommand{\varepsilonv}{\vecs{\varepsilon}}

\newcommand{\thetav}{\vecs{\theta}}
\newcommand{\psiv}{\vecs{\psi}}

\newcommand{\Id}{\vec{I}}
\newcommand{\Jm}{\vec{J}}

\newcommand{\Pm}{\vec{P}}

\newcommand{\Sm}{\vec{S}}

\newcommand{\Xm}{\vec{X}}
\newcommand{\Ym}{\vec{Y}}

\newcommand{\Sigmam}{\vecs{\Sigma}}

\newcommand{\Cc}{{\cal C}}

\newcommand{\Lc}{{\cal L}}

\newcommand{\Nc}{{\cal N}}

\newcommand{\Tc}{{\cal T}}

\newcommand{\Xc}{{\cal X}}
\newcommand{\Yc}{{\cal Y}}

\newcommand{\CC}{\mathbb{C}}
\newcommand{\MM}{\mathbb{M}}

\newcommand{\RR}{\mathbb{R}}

\newcommand{\tp}{^{\mathsf{T}}}

\newcommand{\LB}{\left(}
\newcommand{\RB}{\right)}
\newcommand{\LP}{\left\{}
\newcommand{\RP}{\right\}}
\newcommand{\LSB}{\left[}
\newcommand{\RSB}{\right]}

\renewcommand{\log}[1]{\mathop{\mathrm{log}}\LB #1\RB}
\renewcommand{\exp}[1]{\mathop{\mathrm{exp}}\LB #1\RB}

\newcommand{\EE}{{\mathbb{E}}}

\newcommand\norm[1]{\left\lVert#1\right\rVert}
\newcommand\abs[1]{\left| #1 \right|}

\begin{document}
\title{Model-free Training of End-to-end Communication Systems}
\author{Fayçal~Ait~Aoudia and~Jakob~Hoydis, \IEEEmembership{Senior Member,~IEEE} %
\thanks{
F. Ait Aoudia and J. Hoydis are with Nokia Bell Labs, Paris-Saclay, 91620 Nozay, France (\{faycal.ait\_aoudia, jakob.hoydis\}@nokia-bell-labs.com). Parts of this work have been presented at the ASILOMAR Conference 2018 \cite{Aoudia2018EndtoEndLO}.}
}

\maketitle

\begin{abstract}

The idea of end-to-end learning of communication systems through \gls{NN}-based autoencoders has the shortcoming that it requires a differentiable channel model.
We present in this paper a novel learning algorithm which alleviates this problem.
The algorithm enables training of communication systems with an unknown channel model or with non-differentiable components.
It iterates between training of the receiver using the true gradient, and training of the transmitter using an approximation of the gradient. 
We show that this approach works as well as model-based training for a variety of channels and tasks.
Moreover, we demonstrate the algorithm's practical viability through hardware implementation on \glspl{SDR} where it achieves state-of-the-art performance over a coaxial cable and wireless channel.


\end{abstract}
\glsresetall

\section{Introduction}

\begin{figure}
\centering
  \includegraphics[draft=false,width=0.7\linewidth]{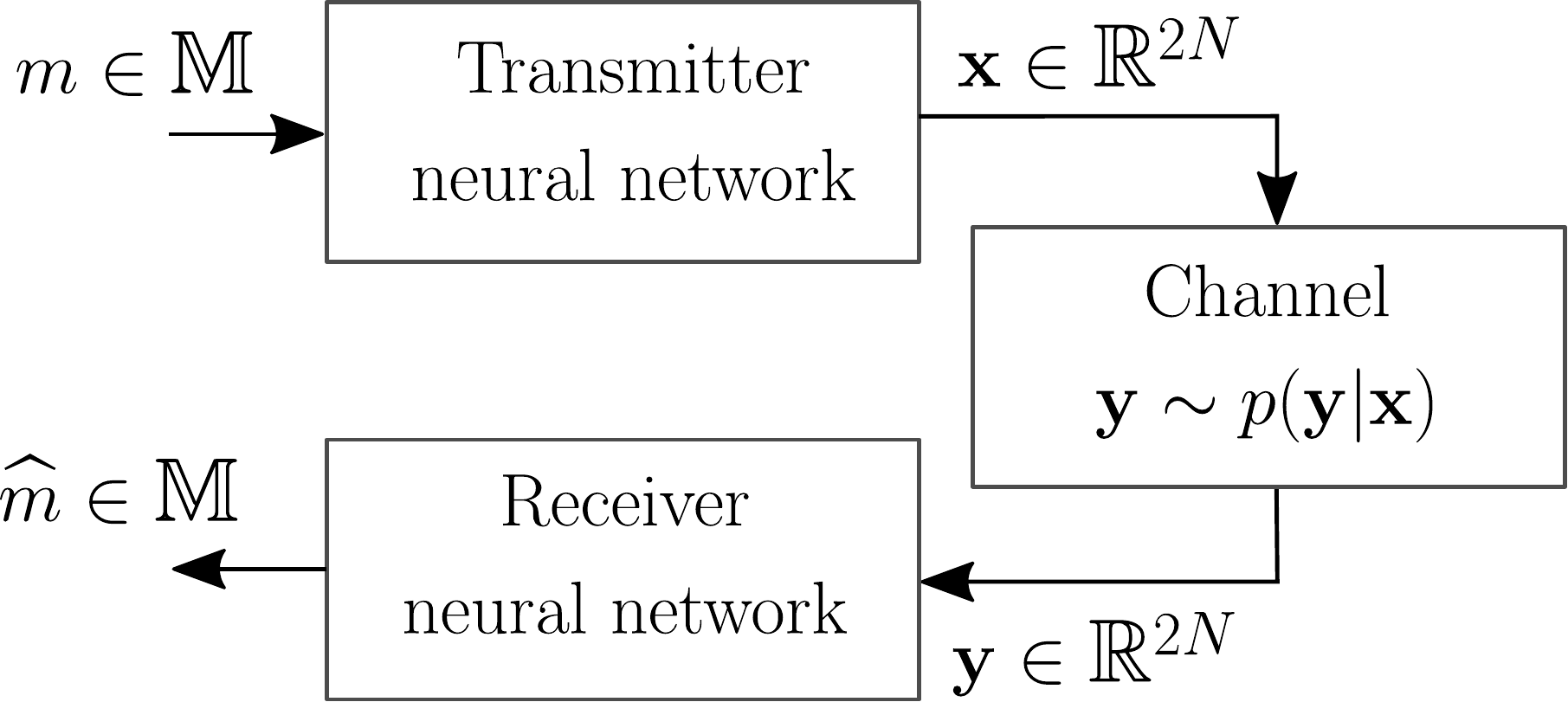}
\caption{An autoencoder-based communication system operating over a channel. End-to-end learning describes the process of jointly optimizing the transmitter and the receiver.}
\label{fig:big_pic}
\end{figure}

End-to-end learning of communication systems is a fascinating novel concept~\cite{8054694} whose goal is to learn full transmitter and receiver implementations  which are optimized for a specific performance metric and channel model.
This can be achieved by representing transmitter and receiver as \glspl{NN}, as illustrated in Fig.~\ref{fig:big_pic}, and by interpreting the whole system as an \emph{autoencoder}~\cite{Goodfellow-et-al-2016-Book} which can be trained in a supervised manner using \gls{SGD}.
Although a theoretically very appealing idea, its biggest drawback hindering practical implementation is that a channel model or, more precisely, the gradient of the instantaneous channel transfer function, must be known~\cite{dorner2017deep}.
For an actual system, this is hardly the case since the channel is generally a black box for which only inputs and outputs can be observed.
Moreover, the channel typically comprises some parts of the transceiver, such as quantization,  which are non-differentiable and hence forbid gradient-based training through backpropagation.


In this work, we provide a method to circumvent the problem of a missing channel gradient.
The key idea is to approximate the loss function gradient \gls{wrt} the transmitter parameters by relaxing the channel input to a random variable.
The proposed approach removes the requirement of channel model knowledge, which implies that the autoencoder can be trained from pure observations alone.
We develop a novel \emph{alternating} algorithm for end-to-end training without channel model knowledge which iterates between two phases: (i) training of the receiver using the true gradient of the loss and (ii) training of the transmitter based on an approximation of the loss function gradient.

We compare the performance of the proposed scheme against that of training with a channel model using usual backpropagation~\cite{8054694} for a variety of channel models and tasks.
Specifically, on \gls{AWGN} and \gls{RBF} channels, both approaches achieve identical performance. On the \gls{AWGN} channel, the learned system outperforms \gls{QPSK} and comes close to Agrell~\cite{Agrell16}, a close-to-optimal solution to the sphere packing problem. On the \gls{RBF} channel, it outperforms both \gls{QPSK} and Agrell.
 Evaluation on a simplified fiber-optic channel model also reveals identical performance of the model-based and model-free  training algorithms, illustrating the universality of the proposed scheme.
Finally, we show experimental results for the first autoencoder-based communication system trained directly over actual channels (wireless and coaxial cable). In contrast to~\cite{dorner2017deep}, the learned system is now able to 
achieve competitive performance \gls{wrt} to a well designed baseline.

\textbf{Notations:}
Boldface upper-case (lower-case) letters denote matrices (column vectors). $\RR$ ($\CC$) is the set of real (complex) numbers. $\Nc(\mv,\Sm)$ is the Gaussian distribution with mean $\mv$ and covariance $\Sm$. For a function $f:\RR^n\mapsto\RR^k$, we denote by $\nabla_{\xv} f(\xv) \in \RR^{n\times k}$ its Jacobian. 

\section{Background}

A point-to-point communication system consists of two nodes that aim to reliably exchange information over a channel.
The channel is a stochastic system whose output $\yv\in\RR^{2N}$ follows a probability distribution conditional on its input $\xv\in\RR^{2N}$, i.e., $\yv \sim p(\yv|\xv)$.
This is equivalent to the complex baseband representation with $\xv,\yv \in\CC^N$.
The transmitter wants to communicate messages $m$ uniformly drawn from a discrete set $\MM = \{1,\dots,M\}$, while the receiver tries to detect the sent messages from the received signals, as illustrated in Fig.~\ref{fig:big_pic}.
Typically, the design of communication systems relies on dividing transmitter and receiver into individual blocks, each performing one task, such as modulation or channel coding.
However, this component-wise approach is not guaranteed to lead to the best possible performance and the attempts to jointly optimize components revealed intractable or too computationally complex~\cite{504941}.
This motivates the use of \gls{ML} to enable optimization of communication systems for end-to-end performance without the need for compartmentalization of transmitter and receiver~\cite{8054694}.

The key idea of autoencoder-based communication systems is to represent the transmitter, channel, and receiver as a single \gls{NN}---the autoencoder---which aims to reproduce its input at its output.
If a differentiable model of the channel is available, it can be implemented as non-trainable layers, between the transmitter and the receiver, and the end-to-end system is trained using \gls{SGD}.
This idea was pioneered in~\cite{8054694} and the first proof-of-concept using off-the-shelf \glspl{SDR} was described in~\cite{dorner2017deep}.
Numerous extensions of the original idea towards channel coding~\cite{kim2018communication}, \gls{OFDM}~\cite{kimOFDM2018, felix2018ofdm, balevi2018one}, \gls{MIMO}~\cite{osheamimo2017}, as well as \gls{SWIPT} \cite{varasteh2018learning} have been made, which all demonstrate the versatility of this approach.
Another line of work considers the autoencoder idea for joint source-channel coding~\cite{farsad2018text, bourtsoulatze2018deep}.

A limitation of the original approach \cite{8054694} is the requirement of a differentiable channel model to train the transmitter, by backpropagating the gradient through the channel layers.
However, in practice, such a channel model is hardly available, and training using inaccurate channel models leads to significant performance loss once the system is deployed~\cite{dorner2017deep}.
A simple work-around proposed in this latter work \rev{consists of} fine-tuning of the receiver based on measured data after initial training on a channel model.
However, with this approach, the transmitter cannot be fine-tuned to the actual channel, resulting in sub-optimal performance.
An alternative approach was proposed in~\cite{ye2018channel,o2018approximating}. The key idea is to learn a differentiable generative model of the channel in the form of a \gls{GAN}, which can then be used to train the autoencoder.
However, it has not yet been not shown that this approach works for practical channels.
The idea of training an \gls{NN}-based transmitter using policy gradient was explored in~\cite{devrieze2018multiagent} for a non-differentiable receiver which treats detection as a clustering problem.
Their approach does not outperform a standard modulation baseline and was only evaluated on the \gls{AWGN} channel.
We have proposed in~\cite{Aoudia2018EndtoEndLO} an alternating algorithm for autoencoder training without channel model.
In the current paper, we provide a theoretical understanding of this algorithm, carry out a  wide set of simulation-based experiments, and report on results from a first hardware implementation.
The follow-up work \cite{Raj2018} to \cite{Aoudia2018EndtoEndLO} explores \gls{SPSA} for approximating the loss function gradient.
However, \rev{as shown in Appendix~\ref{app:spsa}}, this method does not scale well with the number of trainable parameters of the transmitter~\gls{NN}.

\section{Gradient Estimation Without Channel Model} \label{sec:algo}

The key idea behind the approach proposed in this work is to implement transmitter and receiver as two separate parametric functions that are jointly optimized to meet application specific performance requirements. The transmitter maps a message $m$ to channel symbols $\xv$, and is represented by the mapping $f^{(T)}_{\thetav_T}: \MM \to \RR^{2N}$, where $N$ is the number of complex channel uses and $\thetav_T$ is the vector of parameters. The receiver is implemented as $f^{(R)}_{\thetav_R} : \RR^{2N} \to \LP \pv \in \RR_+^M~|~\sum_{i=1}^M p_i = 1 \RP$, where $\thetav_R$ is the vector of parameters and $\pv$ a probability vector over the messages.
We assume that $f^{(T)}_{\thetav_T}$ and $f^{(R)}_{\thetav_R}$ are differentiable \gls{wrt} their parameters, which are adjusted through gradient descent on the loss function
\begin{equation}\label{eq:loss}
    \Lc(\thetav_T, \thetav_R) \triangleq
    \EE_m \left\{ \int l\left(f^{(R)}_{\thetav_R}(\mathbf{y}),m\right) p\left(\mathbf{y}~\big|~f^{(T)}_{\thetav_T}(m)\right) d\mathbf{y} \right\}
\end{equation}
where $\EE_m$ is the expectation taken over the messages $m$, and $l(\mathbf{p}, m) = -\log{p_m}$ is the categorical \gls{CE}.
It is assumed that $l$ is bounded, which is achieved by adding a small positive constant inside the logarithm.
This trick is widely used in practice to ensure numerical stability.
Parameter optimization is performed using gradient descent, or a variant, which at each iteration requires the computation of the loss function gradient
$\nabla_{(\thetav_R,\thetav_T)} \Lc=
\LSB (\nabla_{\thetav_R} \Lc)\tp, 
(\nabla_{\thetav_T} \Lc)\tp\RSB\tp$.


\subsection{Gradient of the receiver}

The gradient of $\Lc$ \gls{wrt} the receiver parameters $\thetav_R$ is
\begin{equation}
    \nabla_{\thetav_R} \Lc =
    \mathbb{E}_{m,\mathbf{y}} \left\{ \nabla_{\thetav_R} l\left(f^{(R)}_{\thetav_R}(\mathbf{y}),m\right) \right\}. \label{eq:rx_grad}
\end{equation}
The exchange of integration and differentiation, performed here and in the rest of this section, requires regularity conditions, discussed, for example, in~\cite{l1995note}.
Note that $p(\yv|\xv)$ does not need to be differentiable.
The gradient in \eqref{eq:rx_grad} can be estimated through sampling as
\begin{equation}
  \underline{\nabla_{\thetav_R} \Lc} \triangleq \frac{1}{S}\sum_{i=1}^{S}  \nabla_{\thetav_R} l\left(f_{\thetav_R}(\mathbf{y}^{(i)}),m^{(i)}\right) \label{eq:rx_grad_approx}
\end{equation}
where $S$ is the \emph{batch size}, i.e., the number of samples used to estimate the loss value, $m^{(i)}$ is the $i$th training sample, and $\yv^{(i)}$ is the corresponding received signal.
This estimator is valid if the training samples are \gls{iid}.
Since computing \eqref{eq:rx_grad_approx} requires only sampling of the channel output, training of the receiver can be performed without knowledge of the actual channel model $p(\yv|\xv)$.

\subsection{Gradient of the transmitter}

Regarding the transmitter, the gradient of $\Lc$ \gls{wrt} $\thetav_T$ is
\begin{multline} \label{eq:tx_grad}
    \nabla_{\thetav_T} \Lc =
    \mathbb{E}_m \Bigg\{ \int l\left(f^{(R)}_{\thetav_R}(\mathbf{y}),m\right) \nabla_{\thetav_T} f^{(T)}_{\thetav_T}(m)\\
     \cdot \nabla_{\mathbf{x}} p \LB \yv|\xv \RB\big|_{\xv = f^{(T)}_{\thetav_T}(m)} d\mathbf{y} \Bigg\}
\end{multline}
where $\nabla_{\thetav_T} f^{(T)}_{\thetav_T}(m)$ is the Jacobian of the transmitter output and $\nabla_{\mathbf{x}} p \LB \yv| \xv\RB\big|_{\xv=f^{(T)}_{\thetav_T}(m)}$ is the gradient of the channel $p(\yv|\xv)$ \gls{wrt} to its input evaluated at $\xv=f^{(T)}_{\thetav_T}(m)$. 
As $p(\yv|\xv)$ is not known, and may not be differentiable, its gradient cannot be calculated and might even be undefined.
\rev{
A workaround is to see the channel input $\xv$ as a random variable that follows a distribution $\pi_{\bar{\xv}} = \delta(\xv - \bar{\xv})$ parametrized by $\bar{\xv}$, with $\delta$ the Dirac distribution, and rewrite the loss~\eqref{eq:loss} as
\begin{multline}
    \Lc(\thetav_T, \thetav_R)
    = \mathbb{E}_m \Bigg\{ \int \pi_{f^{(T)}_{\thetav_T}(m)}(\xv)\\
    \cdot \int l\LB f^{(R)}_{\thetav_R}(\yv),m \RB p \LB \yv|\xv \RB d\yv d\xv \Bigg\}.
\end{multline}}
We then relax $\xv$ to follow a distribution $\hat{\pi}_{\bar{\xv},\sigma}$ parametrized by $\bar{\xv}$ and its standard deviation $\sigma > 0$.
We denote by $\widehat{\Lc}$ the associated loss 
\begin{multline} \label{eq:loss_hat}
    \widehat{\Lc}(\thetav_T, \thetav_R) \triangleq \EE_m \Big\{ \int \hat{\pi}_{f^{(T)}_{\thetav_T}(m), \sigma}(\xv)\\
    \cdot \int l \LB f^{(R)}_{\thetav_R}(\yv),m \RB p \LB \yv|\xv \RB d\yv d\xv \Bigg\}
\end{multline}
with gradient
\begin{align} 
    &\nabla_{\thetav_T}\widehat{\Lc} \nonumber\\
    & = \EE_m \Bigg\{ \int \nabla_{\thetav_T} \hat{\pi}_{f^{(T)}_{\thetav_T}(m),\sigma}(\xv) \int l \LB f^{(R)}_{\thetav_R}(\yv),m \RB p(\yv|\xv) d\yv d\xv \Bigg\} \nonumber\\
    &\begin{aligned} = \EE_m \Bigg\{ \int  \nabla_{\thetav_T} f^{(T)}_{\thetav_T}(m) \nabla_{\bar{\xv}} \hat{\pi}_{\bar{\xv},\sigma}(\xv)\big|_{\bar{\xv}=f^{(T)}_{\thetav_T}(m)}\\
    \cdot \int l \LB f^{(R)}_{\thetav_R}(\yv),m \RB p(\yv|\xv) d\yv d\xv \Bigg\}
    \end{aligned}\nonumber\\
    &\begin{aligned} = \EE_m \Bigg\{ \int  \nabla_{\thetav_T} f^{(T)}_{\thetav_T}(m) \hat{\pi}_{f^{(T)}_{\thetav_T}(m),\sigma}(\xv) \nabla_{\bar{\xv}} \log{\hat{\pi}_{\bar{\xv},\sigma}(\xv)}\big|_{\bar{\xv}=f^{(T)}_{\thetav_T}(m)}\\
    \cdot \int l \LB f^{(R)}_{\thetav_R}(\yv),m \RB p(\yv|\xv) d\yv d\xv \Bigg\}
    \end{aligned}\nonumber\\
    &\begin{aligned} = \EE_{m,\xv,\yv} \Bigg\{ l \LB f^{(R)}_{\thetav_R}(\yv),m \RB \nabla_{\thetav_T} f^{(T)}_{\thetav_T}(m)\\
    \cdot \nabla_{\bar{\xv}} \log{\hat{\pi}_{\bar{\xv},\sigma}(\xv)}\big|_{\bar{\xv}=f^{(T)}_{\thetav_T}(m)} \Bigg\}
    \end{aligned}
    \label{eq:L_hat_grad}
\end{align}
where the second equality leverages the chain rule, and the third equality uses the log-trick $\nabla_{\xv} \log{g(\xv)} = \frac{\nabla_{\xv}g(\xv)}{g(\xv)}$.
Computing $\nabla_{\thetav_T}\widehat{\Lc}$ does not require differentiability of $p(\yv|\xv)$, and it can be simply estimated through sampling of the channel distribution:
\begin{multline}
  \underline{\nabla_{\thetav_T}\widehat{\Lc}} \triangleq \frac{1}{S}\sum_{i=1}^{S} l\left(f^{(R)}_{\thetav_R}(\mathbf{y}^{(i)}),m^{(i)}\right) \nabla_{\thetav_T} f^{(T)}_{\thetav_T}(m^{(i)})\\
  \cdot \nabla_{\bar{\xv}} \log{\hat{\pi}_{\bar{\xv},\sigma}(\xv^{(i)})}\Big|_{\bar{\xv}=f^{(T)}_{\thetav_T}(m^{(i)})}. \label{eq:L_hat_grad_approx}
\end{multline}

The gradient estimate in \eqref{eq:L_hat_grad_approx} allows optimization of the transmitter \gls{wrt} $\widehat{\Lc}$ without knowledge of the underlying channel model and has multiple interpretations. From the viewpoint of \gls{RL} (as done in our previous work~\cite{Aoudia2018EndtoEndLO}), the transmitter is an \emph{agent} in \emph{state} $m$, which interacts with an \emph{environment} formed by the \{channel, receiver\} system, taking \emph{actions} $\xv$ following the \emph{policy} $\hat{\pi}_{\bar{\xv},\sigma}$, and receiving \emph{penalties} $l$ conditional to the chosen action and current state.
The goal of the \gls{RL} agent is to optimize its policy which indicates which action to take given the current state in order to minimize the average received penalty.
Note that \gls{RL} is widely used to optimize agents when no model of the environment is available and/or when the penalty function is not differentiable.
From this viewpoint, relaxation of the transmitter outputs to a random distribution can be seen as relaxing a deterministic policy, which maps each input message $m$ to channel symbols $\xv$ in a deterministic way, to a stochastic policy which enables \emph{exploration} of the set of possible actions.

A second interpretation of \eqref{eq:L_hat_grad_approx} and an understanding of the relationship between $\nabla_{\thetav_T}\widehat{\Lc}$ and $\nabla_{\thetav_T}\Lc$, when the later is defined, can be obtained from studying the difference
\begin{align}\notag
  & \nabla_{\thetav_T} \Lc - \nabla_{\thetav_T} \widehat{\Lc}\\
  &\begin{aligned} = \EE_m \Bigg\{ \int l \LB f^{(R)}_{\thetav_R}(\yv), m \RB \nabla_{\thetav_T} f^{(T)}_{\thetav_T}(m)
  \bigg[ \nabla_{\xv} p \LB \yv|\xv \RB\big|_{\xv=f^{(T)}_{\thetav_T}(m)}\\
  - \EE_{\xv} \LP p \LB \yv | \xv \RB \nabla_{\bar{\xv}} \log{\hat{\pi}_{\bar{\xv},\sigma}(\xv)}\big|_{\bar{\xv}=f^{(T)}_{\thetav_T}(m)} \RP \bigg] d\yv \Bigg\}.
  \end{aligned} \label{eq:gradient_diff}
\end{align}
One can see that the difference between the two gradients goes to zero when the true channel gradient $\nabla_{\xv}p(\yv|\xv)$ is well approximated by $\EE_{\xv} \LP p \LB \yv | \xv \RB \nabla_{\bar{\xv}} \log{\hat{\pi}_{\bar{\xv},\sigma}(\xv)}\big|_{\bar{\xv}=f^{(T)}_{\thetav_T}(m)} \RP$.
Therefore, the relaxation of the transmitter output to a random variable can be seen as a way to estimate and approximate the unknown channel gradient.
This intuition is comforted by the following theorem, which holds under some requirements on $\hat{\pi}_{\bar{\xv},\sigma}$ and the channel distribution.
\begin{thm}
\label{tm:lim_sigma}
Assume that $\hat{\pi}_{\bar{\xv},\sigma}$ and $p(\yv|\xv)$ satisfy conditions given in the \rev{Appendix~\ref{app:proof}}, then
\begin{equation}
    \lim_{\sigma \to 0} \nabla_{\thetav_T} \widehat{\Lc}(\thetav_T, \thetav_R) = \nabla_{\thetav_T} \Lc(\thetav_T, \thetav_R).
\end{equation}
\end{thm}
\begin{IEEEproof}
See the \rev{Appendix~\ref{app:proof}}.
\end{IEEEproof}

Theorem~\ref{tm:lim_sigma} states that, under some conditions, the true loss function gradient \gls{wrt} transmitter parameters $\nabla_{\thetav_T} \Lc$ can be approximated with arbitrarily small error by the substitute loss gradient $\nabla_{\thetav_T} \widehat{\Lc}$.
A similar results was provided in~\cite[Theorem~2]{silver2014deterministic} in the context of \gls{RL}.
However, to approximate the true gradient, a function, denoted by $Q$, which provides the expected loss given the state and action is needed~\cite[Theorem~1]{silver2014deterministic}.
In practice, this function is approximated using an additional \gls{NN} that is trained together with the \gls{NN} implementing the policy.
Although this approach might work well when the channel and receiver are fixed, it is unpractical in our case. Since the receiver is jointly trained with the transmitter, one would need to re-train the \gls{NN} approximating the $Q$ function at every update of receiver parameters.
Moreover, the \gls{NN} approximating the $Q$ function needs to be \emph{compatible}, i.e., it must satisfy some conditions which are hard to achieve in practice (see \cite[Theorem~3]{silver2014deterministic} and the following discussion).

We have so far provided a theoretical motivation for the gradient approximation in \eqref{eq:L_hat_grad}, by showing that $\nabla_{\thetav_T} \widehat{\Lc}$ approximates $\nabla_{\thetav_T} \Lc$ under some reasonable conditions.
In particular, we assumed that $p(\yv|\xv)$ is differentiable \gls{wrt} $\xv$.
However, the gradient estimators in \eqref{eq:rx_grad_approx} and \eqref{eq:L_hat_grad_approx} do not require differentiability of $p(\yv|\xv)$
and can even be computed in cases where $p(\yv|\xv)$ is not differentiable. 
Providing theoretical guarantees in this case is an open problem.


\section{Training End-to-End Communication Systems} 

We now present an \emph{alternating} training algorithm which works for any pair of differentiable parametric functions $f^{(T)}_{\thetav_T}, f^{(R)}_{\thetav_R}$.
We choose to implement them here as \glspl{NN}.
Since this algorithm does not require a channel model, we will refer to it also as the \emph{model-free} training method.
\emph{Model-aware} training is used to refer to training with channel model knowledge.
With model-aware training, transmitter, channel model, and receiver form a single \gls{NN} which is trained \rev{using the usual backpropagation} as in~\cite{8054694}.
\rev{However, this approach requires the use (and knowledge) of a differentiable channel model.}


\subsection{Generic transmitter and receiver architectures} \label{sec:gen_arch}

\begin{figure}
    \centering
  \begin{subfigure}{0.40\linewidth}
    \centering
    \includegraphics[width=0.76\linewidth]{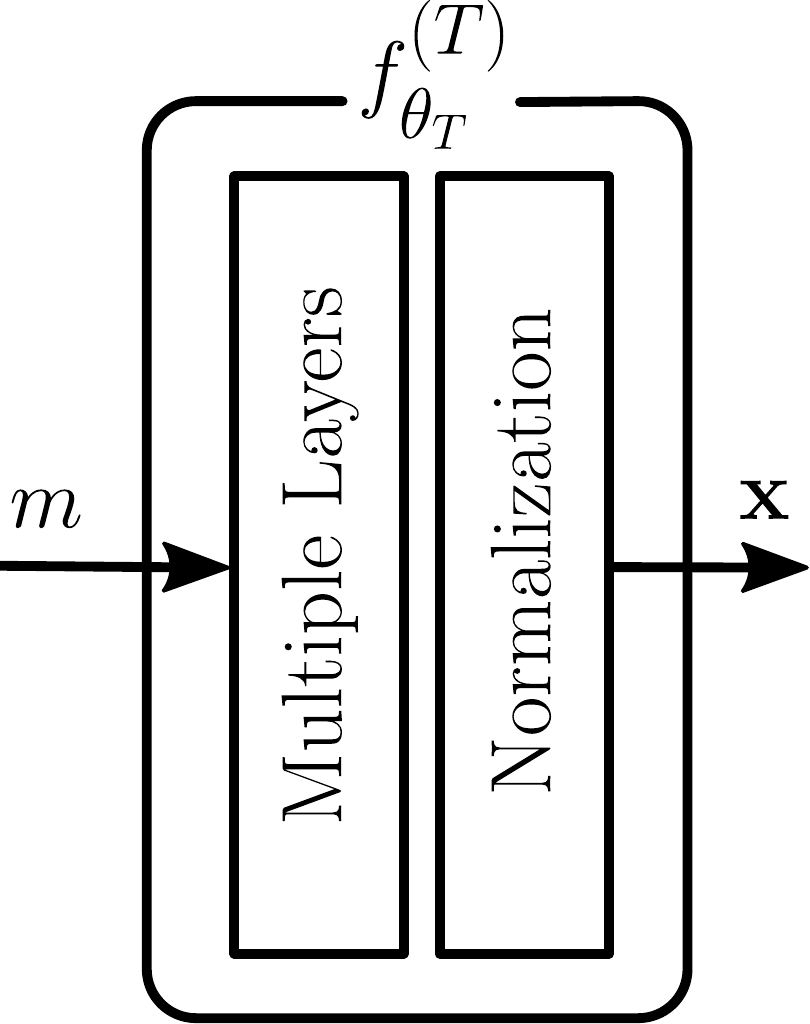}
    \subcaption{Generic transmitter \gls{NN}}
    \label{fig:encoder_arch}
  \end{subfigure}\hspace{5mm}
  \begin{subfigure}{0.45\linewidth}
    \centering
    \includegraphics[width=\linewidth]{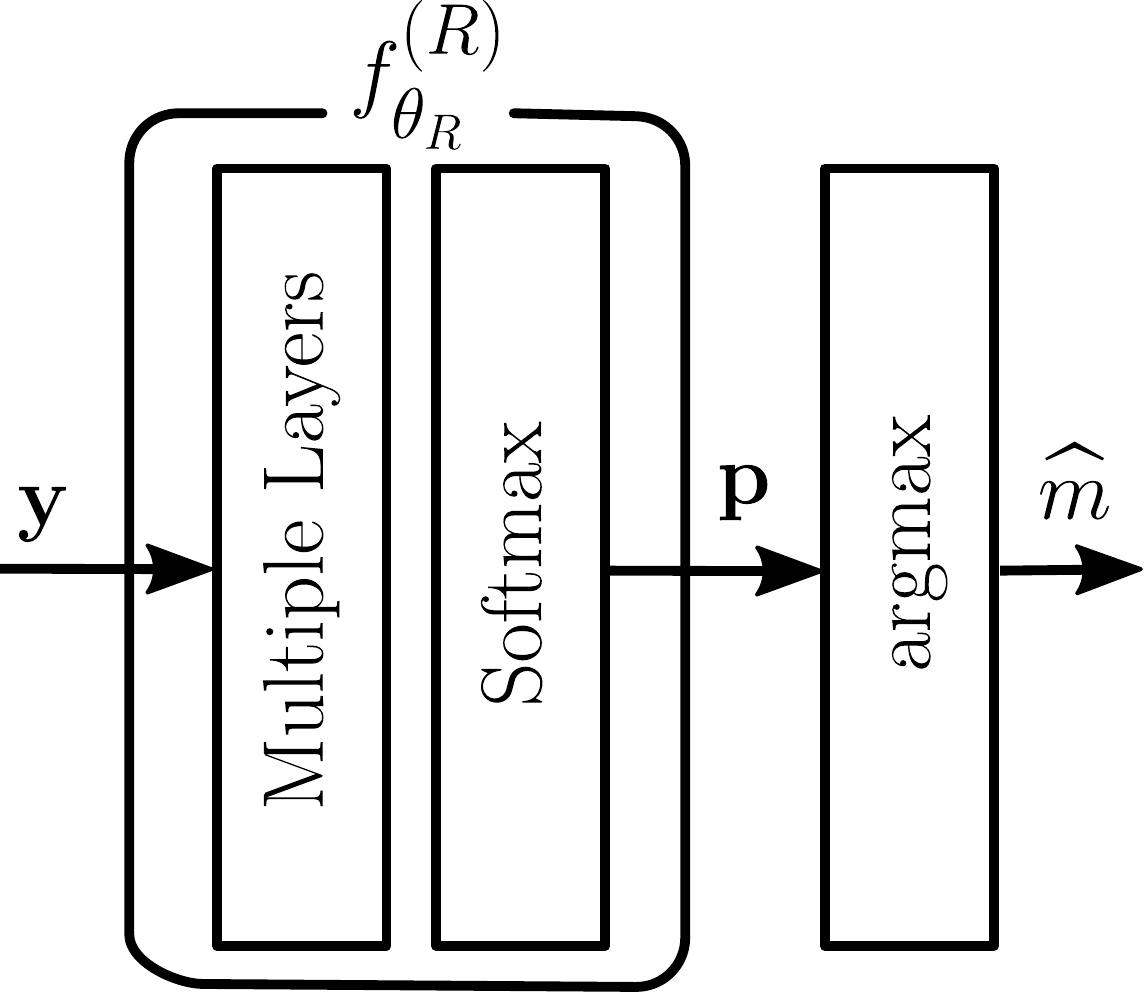}
    \subcaption{Generic receiver \gls{NN}}
    \label{fig:decoder_arch}
  \end{subfigure}
  \caption{Generic architectures of the transmitter and receiver}
  \label{fig:nn_gen_arch}
\end{figure}

The architectures of the transmitter and the receiver can take multiple forms.
However, in the context of communication systems, the transmitter must ensure the fulfillment of power and possibly other hardware-dependent constraints.
Therefore, its last layer performs normalization which guarantees, e.g., that the average energy per symbol or per message is one.
All non-differentiable operations on the transmitter output, e.g., quantization, can be assumed to be part of the channel.
A generic architecture of the transmitter is shown in Fig.~\ref{fig:encoder_arch}.
The receiver task is to detect the sent message from the received signal.
The received signal is fed to a succession of layers that can be arbitrarily chosen.
The last layer of $f_{\thetav_R}^{(R)}$ is a softmax layer to ensure that the output activations form a probability vector over $\MM$~\cite{Goodfellow-et-al-2016-Book}. Finally, the index with the highest probability is chosen as a hard decision on the sent message.
A generic architecture of the receiver is shown in Fig.~\ref{fig:decoder_arch}.
As for the transmitter, one can model non-differentiable receiver components as parts of the channel.

\subsection{Training process overview}
\label{sec:training}


The key idea of the proposed training algorithm is to alternately train the receiver using the true gradient (\ref{eq:rx_grad}), and the transmitter using the gradient approximation (\ref{eq:L_hat_grad}).
When training the receiver (transmitter), the parameters of transmitter (receiver) are kept fixed.
Therefore, an iteration of the training algorithm is made of two phases, one for the receiver (Sec.~\ref{subsec:rxtrain}), and one for the transmitter (Sec.~\ref{subsec:txtrain}), as shown in Algorithm~\ref{alg:training}.
This training process is carried out until a stop criterion is satisfied (e.g., a fixed number of iterations, a fixed number of iterations during which the loss has not decreased, etc.).

\begin{algorithm}
\caption{Alternating training algorithm}
\label{alg:training}
\begin{algorithmic}[1]
\Repeat
  \State \Call{TrainReceiver}{\null}
  \State \Call{TrainTransmitter}{\null}
\Until{Stop criterion is met}
\end{algorithmic}
\end{algorithm}

It is assumed that transmitter and receiver have access to a sequence of training examples $m_{\Tc}^{(i)},~i=1,2,\dots$.
This can be achieved through pseudorandom number generators initialized with the same seed.
Training of transmitter and receiver is done by gradient descent or, more precisely, \gls{SGD}~\cite{Goodfellow-et-al-2016-Book} or one of its numerous variants.
The true gradient in \eqref{eq:rx_grad} for the receiver is estimated by \eqref{eq:rx_grad_approx}, while the gradient approximation \eqref{eq:L_hat_grad} used to train the transmitter is estimated by \eqref{eq:L_hat_grad_approx}.
Both training phases are described next.
For convenience, we adopt a matrix notation, commonly used in \gls{ML}, where each row of a matrix corresponds to a training example of a minibatch.

\subsection{Receiver training}\label{subsec:rxtrain}

\begin{figure*}
  \centering
  \begin{subfigure}{0.42\linewidth}
    \centering
    \includegraphics[width=\linewidth]{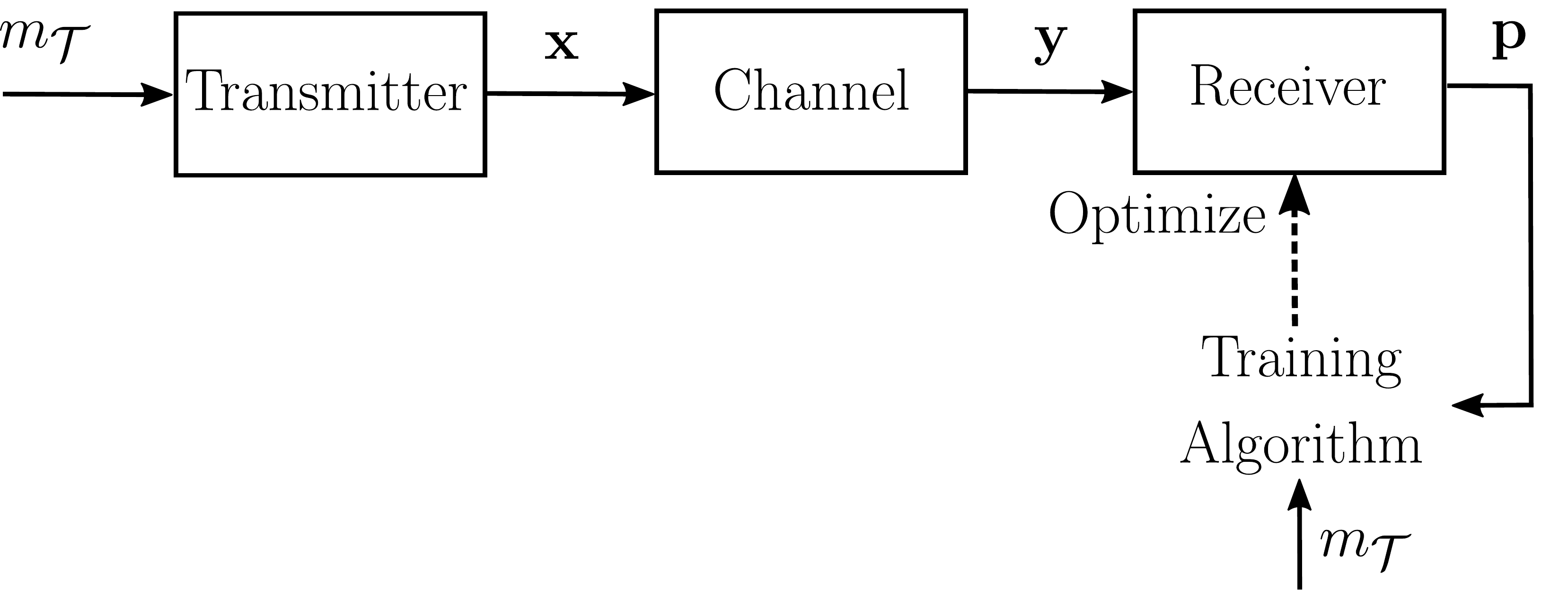}
    \subcaption{Receiver training}
    \label{fig:train_decoder} 
  \end{subfigure}\qquad
  \begin{subfigure}{0.47\linewidth}
    \centering
    \includegraphics[width=\linewidth]{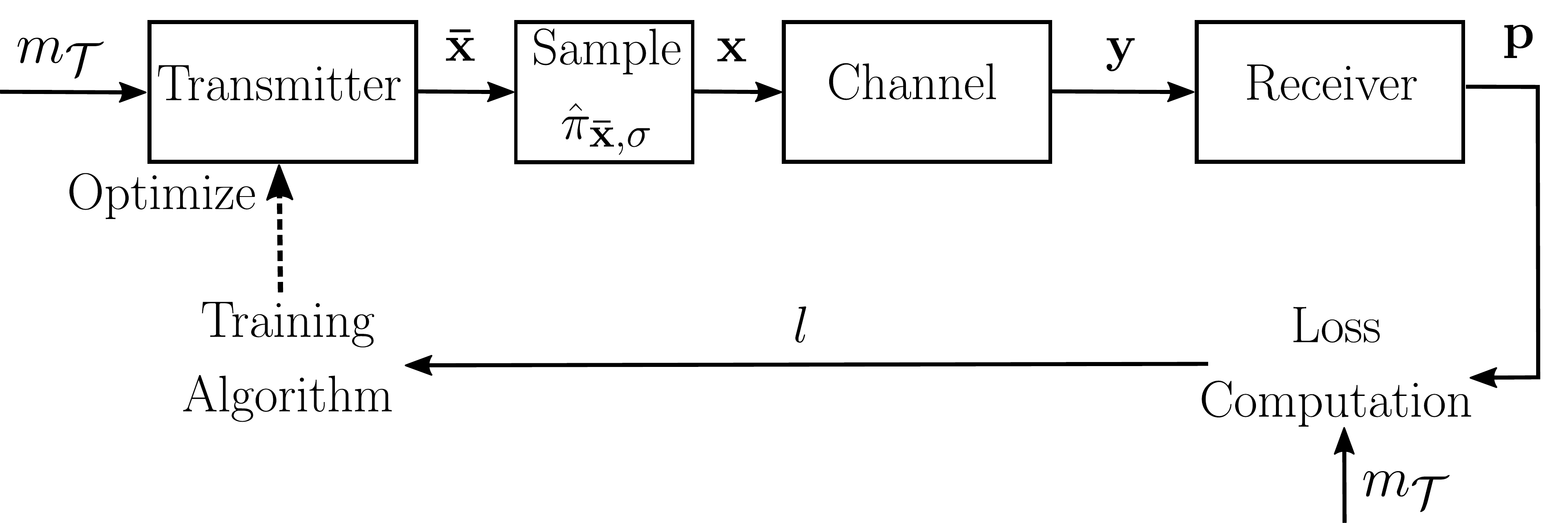}
    \subcaption{Transmitter training}
    \label{fig:train_encoder}
  \end{subfigure}
  \caption{The two phases of a training iteration: receiver training and transmitter training}
\end{figure*}

The receiver training process is illustrated in Fig.~\ref{fig:train_decoder}, and the pseudocode is given in Algorithm~\ref{alg:training_rx}.
First, the transmitter generates a minibatch of messages $\mv_{\Tc}$ of size $B_R$.
Then, each message is encoded into $2N$ ($N$) real (complex) channel symbols, where $\Xm$ is the $B_R$-by-$2N$ matrix containing the symbols (lines~\ref{lst:td_data_send_s}--\ref{lst:td_data_send_e}).
\rev{No relaxation of the transmitter output is performed, as it is not required to approximate the receiver gradient.}
The encoded minibatch is then sent over the channel.
The receiver obtains the altered symbols $\Ym$, from which it generates for each training example a probability vector over $\MM$, \rev{stacked into a matrix $\Pm$ of size $B_R-$by$-M$} (lines~\ref{lst:td_data_recv_s}--\ref{lst:td_data_recv_e}).
Finally, an optimization step is performed using the estimated gradient (\ref{eq:rx_grad_approx}) (line~\ref{lst:td_sgd_e}).

\begin{algorithm}
\caption{Training of the receiver}
\label{alg:training_rx}
\begin{algorithmic}[1]
\Function{TrainReceiver}{\null}
  \Repeat
    \State \LineComment{Transmitter:}
    \State $\mv_{\Tc} \gets \Call{TrainingSource}{B_R}$\label{lst:td_data_send_s}
    \State $\Xm \gets f^{(T)}_{\thetav_T}(\mv_{\Tc})$ \label{lst:td_data_send_e}
    \State $\Call{Send}{\Xm}$  \label{lst:td_data_recv_s}
    \State \LineComment{Receiver:}
    \State $\Ym \gets \Call{Receive}{\null}$ 
    \State $\Pm \gets f^{(R)}_{\thetav_R}(\Ym)$ \label{lst:td_data_recv_e}
    \State $\Call{SGD}{f^{(R)}_{\thetav_R}, \mv_{\Tc}, \Pm}$ \label{lst:td_sgd_e}
  \Until{Stop criterion is met}
\EndFunction
\end{algorithmic}
\end{algorithm}

\subsection{Transmitter training}\label{subsec:txtrain}


The pseudocode of the transmitter training process is shown in Algorithm~\ref{alg:training_tx}, and illustrated in Fig.~\ref{fig:train_encoder}.
First, a minibatch $\mv_{\Tc}$ of size $B_T$ is generated, and each training example is encoded into $2N$ channel symbols to form the $B_T$-by-$2N$ matrix $\overline{\Xm}$ (lines~\ref{lst:te_data_enc_s}--\ref{lst:te_data_enc_e}).
Then, relaxation of the transmitter output to a random variable is made by sampling the distribution $\hat{\pi}_{\overline{\Xm}, \sigma}$, and the so-obtained samples are stacked to form the $B_T$-by-$2N$ matrix $\Xm$ (line~\ref{lst:te_relax}).
These symbols are then sent over the channel.
The receiver obtains the altered symbols $\Ym$ and generates for each training example a probability vector over $\MM$, stacked to form the matrix $\Pm$ (lines~\ref{lst:te_data_recv_s}--\ref{lst:te_data_recv_e}).
Per-example losses $\lv\in\RR^{B_T}$ are then computed based on these probability vectors and the sent messages $\mv_{\Tc}$ (line~\ref{lst:te_lg_e}).
Next, the per-example losses are sent to the transmitter over a reliable feedback link, present only during training (lines~\ref{lst:te_loss_send_s}--\ref{lst:te_loss_send_e}).
Finally, an optimization step is performed, using \gls{SGD} or a variant, where the loss gradient is estimated by (\ref{eq:L_hat_grad_approx}).

\begin{algorithm}
\caption{Training of the transmitter}
\label{alg:training_tx}
\begin{algorithmic}[1]
\Function{TrainTransmitter}{\null}
  \Repeat
    \State \LineComment{Transmitter:}
    \State $\mv_{\Tc} \gets \Call{TrainingSource}{B_T}$ \label{lst:te_data_enc_s}
    \State $\overline{\Xm} \gets f^{(T)}_{\thetav_T}(\mv_{\Tc})$ \label{lst:te_data_enc_e}
    \State $\Xm \gets \Call{Sample}{\hat{\pi}_{\overline{\Xm}, \sigma}}$ \label{lst:te_relax}
    \State $\Call{Send}{\Xm}$
    \State \LineComment{Receiver:}
    \State $\Ym \gets \Call{Receive}{\null}$ \label{lst:te_data_recv_s}
    \State $\Pm \gets f^{(R)}_{\thetav_R}(\Ym)$ \label{lst:te_data_recv_e}
    \State $\lv \gets \Call{PerExampleLosses}{\mv_{\Tc}, \Pm}$ \label{lst:te_lg_e}
    \State $\Call{SendPerExampleLosses}{l}$ \label{lst:te_loss_send_s}
    \State \LineComment{Transmitter:}
    \State $\lv \gets \Call{ReceivePerExampleLosses}{\null}$ \label{lst:te_loss_send_e}
    \State $\Call{SGD}{f^{(T)}_{\thetav_T}, \lv}$ \label{lst:te_opt}
  \Until{Stop criterion is met}
\EndFunction
\end{algorithmic}
\end{algorithm}



\section{Evaluation by Simulations}

In this section, we evaluate the performance of the model-free algorithm by simulations.
Relaxation is achieved by adding a zero-mean Gaussian vector $\wv \thicksim \Nc(\zerov, \sigma^2\Id)$ to the transmitter output:
\begin{equation}
    \xv = \sqrt{1 - \sigma^2}f^{(T)}_{\mathbf{\theta_T}}(m) + \wv
\end{equation}
where scaling is done to ensure conservation of the average energy, and $\sigma$ must be chosen in the range $(0,1)$.
From Theorem~\ref{tm:lim_sigma}, one may want to set $\sigma$ to arbitrary small values, as it should lead to a better approximation of the true loss gradient.
However, setting $\sigma$ to small values increases the variance of the estimator (\ref{eq:L_hat_grad_approx}), which leads to slow convergence.
This unwanted effect was experimentally observed, and can be shown analytically in simple settings.
The variance increase due to small values of $\sigma$ can be compensated for by bigger batch sizes, at the cost of a higher computational demand.
Therefore, $\sigma$ controls a tradeoff between the accuracy of the loss function gradient approximation and the estimator variance.
From the viewpoint of \gls{RL}, one can see this also as an interesting tradeoff between the power used for exploration and communication, which would merit an independent study.

The \gls{SNR} is defined as
\begin{equation}
  \text{SNR} = \frac{\EE \LSB \frac1N \lVert\xv \rVert^2_2 \RSB }{2 \sigma_n^2} = \frac{1}{2\sigma_n^2}
\end{equation}
where $2\sigma_n^2$ is the variance per complex baseband noise symbol.
The last equality follows from the normalization step performed by the transmitter, which ensures that $\EE \LSB \frac1N \lVert\xv \rVert^2_2 \RSB$=1.
Normalization is done by scaling the symbols forming a batch.
Note that this is only an approximation for small minibatches.
\rev{Training of the communication systems was done using the Adam~\cite{Kingma15} optimizer, and with a fixed number of iterations which were chosen experimentally.
A training iteration of the alternating algorithm consisted of ten gradient descent steps on the receiver followed by ten gradient descent steps on the transmitter.
The proposed scheme was implemented using the TensorFlow framework~\cite{tensorflow2015-whitepaper}, and evaluated on \gls{AWGN} and \gls{RBF} channels.}
For these channels, the corresponding probability distributions $p(\yv|\xv)$ are respectively $\Nc\LB\yv; \xv, \sigma_n^2\Id\RB$ and $\Nc\LB\yv; \zerov, \frac{1}{2}\LB\xv\xv\tp - \Jm\xv\xv\tp\Jm\RB + \sigma_n^2\Id\RB$, where $\xv$ and $\yv$ are $2N$-dimensional vectors whose components are the $N$ real followed by the $N$ imaginary parts of the $N$ complex baseband symbols, $\Nc(\cdot; \mv, \Sigmam)$ is the probability density function of the multivariate normal distribution with mean $\mv$ and covariance matrix $\Sigmam$, and $\Jm$ is the matrix defined as $
\Jm = 
\begin{bmatrix}
    \zerov  &   -\Id\\
    \Id     & \zerov
\end{bmatrix}
$ where $\zerov$ is the zero square matrix of size $N$ and $\Id$ is the identity matrix of size $N$.

\subsection{Transmitter and receiver architectures}\label{sec:archi}

We implement transmitter and receiver as feedforward \glspl{NN} that leverage only dense layers.
\rev{The messages are fed to the transmitter using the one-hot representation, i.e., each message $m \in \MM$ is encoded as a vector of dimension $M$ which takes as values only zeros, except for the $m$th entry which takes as value one.
The transmitter consists of a dense layer of $M$ units with ELU activation functions~\cite{Goodfellow-et-al-2016-Book}, followed by a dense layer of $2N$ units with linear activations.}
This layer outputs the $2N$ channel symbols, that are finally normalized as shown in Fig.~\ref{fig:encoder_arch}.

Regarding the receiver, the last layer is a dense layer of $M$ units with softmax activations which outputs a probability distribution over $\MM$, as shown in Fig.~\ref{fig:decoder_arch}.
For the \gls{AWGN} channel, a single dense layer with $M$ units and ReLu activation function~\cite{Goodfellow-et-al-2016-Book} was used as hidden layer.
This hidden layer combined with the softmax layer form a \emph{discriminative network}, as its task is to separate the received symbols and assign labels to them, corresponding to guesses of the transmitted messages.
\begin{figure}
    \centering
    \includegraphics[width=0.9\linewidth]{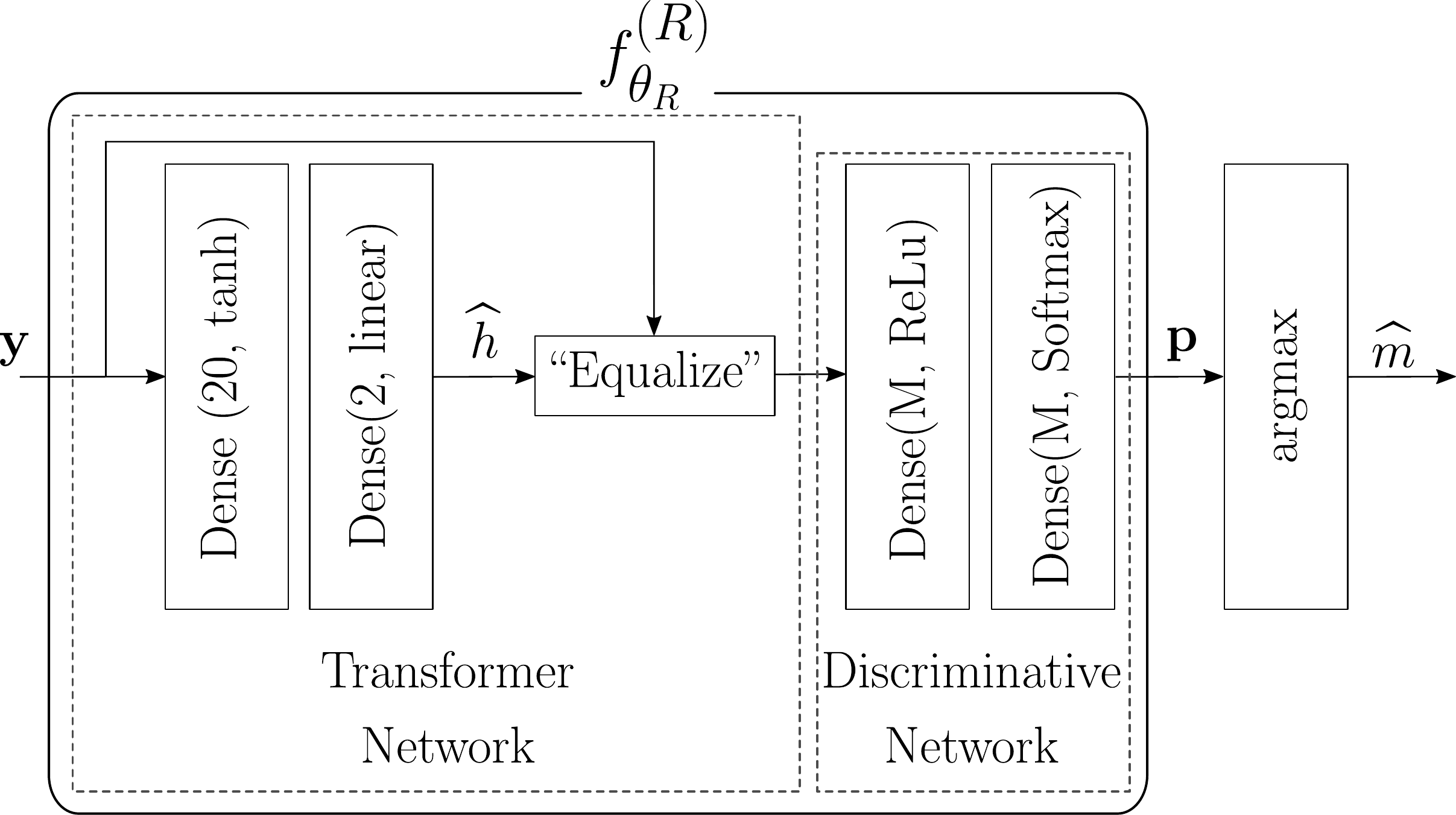}
    \caption{Receiver architecture used for \gls{RBF} channels with no prior equalization}
    \label{fig:decoder_rbf}
\end{figure}
For the \gls{RBF} channel, using only the discriminative network led to poor performance and, therefore, an architecture which incorporates some knowledge about the channel behavior was used \cite[Sec.~III C.]{8054694}.
It is well known in \gls{ML} that incorporating expert knowledge through the \gls{NN} architecture can heavily improve the performance of learning systems.
Accordingly, a \emph{transformer network}, which first transforms the received signal, is added before the discriminative network, as shown in Fig.~\ref{fig:decoder_rbf}.
The transformer network was designed with the intuition that the two first hidden layers produce a value interpreted as an estimate $\widehat{\hv} \in \RR^2$ of the channel response $\hv \in \RR^2$.
As an example, in the case of the \gls{RBF} channel, the channel response is such that $\hv \thicksim \Nc(\zerov, \frac{1}{2}\Id)$.
\rev{
The received signal is then ``equalized'' using the estimated channel response by computing the product of $
\frac{1}{\norm{\widehat{\hv}}^2_2}
\begin{bmatrix}
    \widehat{h_1} \Id   &   \widehat{h_2} \Id\\
    -\widehat{h_2} \Id  &   \widehat{h_1} \Id
\end{bmatrix}
$ and $\yv$, and we refer to the so-obtained signal as the \emph{transformed signal}.
}

\subsection{Evaluation on AWGN and RBF channels}

The size of $\MM$ was set to $M = 256$, corresponding to messages of length 8 bits.
Comparison is done to the \gls{QPSK} and Agrell~\cite{Agrell16} modulation schemes.
For \gls{QPSK}, a message is transmitted by independently sending $2N = 8$ symbols, corresponding to $N = 4$ complex channel uses, each modulating 2 bits of information.
Agrell is a subset of the E8 lattice designed by numerical optimization to approximately solve the sphere packing problem for $M = 256$ in eight dimensions (corresponding to four channel uses).
Model-aware training is also considered, which has been shown to achieve performance close to the best baselines in some scenarios~\cite{8054694}.

Training was done with an \gls{SNR} set to $10\:$dB ($20\:$dB) for the \gls{AWGN} (\gls{RBF}) channel.
For the alternating training algorithm, we used $\sigma = 0.15$.
Using \gls{QPSK} and Agrell with the \gls{RBF} channel, an additional pilot symbol was used to perform explicit equalization.
Regarding the model-aware and model-free schemes with the \gls{RBF} channel, two approaches were considered.
In the first one, the receiver was made of a transformer and a discriminative network.
No pilots were used, $N$ was set to 5 for fairness, and the received symbols were directly fed to the receiver \gls{NN}.
In the second approach, $N$ was set to 4, and a pilot symbol was used to estimate the channel and equalize the received symbols before feeding them to the receiver \gls{NN}, which only consisted of a discriminative network, similarly to the \gls{AWGN} case. Table~\ref{tab:comp_schemes} summarizes the most important parameters of the evaluated schemes.

\begin{table}
\centering
\caption{Evaluated schemes\label{tab:comp_schemes}}
\begin{tabular}{|c|c|c|c|c|}
\hline
Schemes & Pilot & $N$ & \makecell{Block\\Length} & \makecell{\gls{NN} Receiver\\Architecture}\\
\hline
\multicolumn{5}{|c|}{AWGN}\\
\hline
QPSK & None & 4 & 4 & (Not applicable)\\
\hline
Agrell16 & None & 4 & 4 & (Not applicable)\\
\hline
Model-aware/Model-free & None & 4 & 4 & \makecell{Discriminative\\Network}\\
\hline
\multicolumn{5}{|c|}{RBF}\\
\hline
QPSK & 1 & 4 & 5 & \makecell{(Not applicable)}\\
\hline
Agrell16 & 1 & 4 & 5 & \makecell{(Not applicable)}\\
\hline
\makecell{Model-aware/\\Model-free Equalized} & 1 & 4 & 5 & \makecell{Discriminative\\Network}\\
\hline
\makecell{Model-aware/\\Model-free Not Equalized} & None & 5 & 5 & \makecell{Transformer\\ \& Discriminative\\Networks} \\
\hline
\end{tabular}
\end{table}

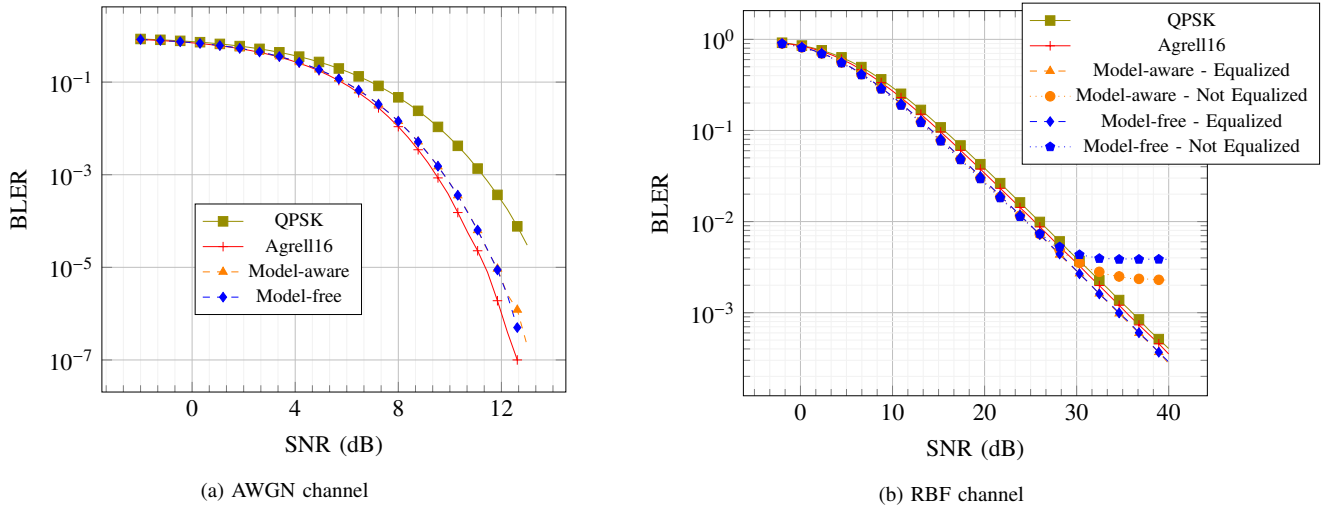
\begin{figure*}
    \centering
    \begin{subfigure}{0.45\textwidth}
    \centering
    \begin{tikzpicture}[scale=0.9]
      \begin{axis}[
        ymode=log,
        grid=both,
        grid style={line width=.4pt, draw=gray!10},
        mark options={solid,scale=1.0},
        major grid style={line width=.2pt,draw=gray!50},
        minor tick num=5,
        xlabel={SNR (dB)},
        ylabel={BLER},
        legend style={at={(0.2, 0.2), font=\footnotesize},anchor=south west},
        xtick={-4, 0, 4, 8, 12, 16},
        mark repeat={2}
      ]
        \addplot[olive, mark=square*] table [x=snr, y=qpsk, col sep=comma] {figs/awgn_sim_bler_vs_snr.csv};
        \addplot[red, mark=+] table [x=snr, y=agrell, col sep=comma] {figs/awgn_sim_bler_vs_snr.csv};
        \addplot[dashed, orange, mark=triangle*] table [x=snr, y=sl, col sep=comma] {figs/awgn_sim_bler_vs_snr.csv};
        \addplot[dashed, blue, mark=diamond*] table [x=snr, y=rl, col sep=comma] {figs/awgn_sim_bler_vs_snr.csv};

      \addlegendentry{QPSK}
      \addlegendentry{Agrell16}
      \addlegendentry{Model-aware}
      \addlegendentry{Model-free}

        \end{axis}

        \end{tikzpicture}
        \caption{AWGN channel}
        \label{fig:awgn_err_vs_snr}
    \end{subfigure}\qquad
    \begin{subfigure}{0.45\textwidth}
    \centering
    \begin{tikzpicture}[scale=0.9]
      \begin{axis}[
        ymode=log,
        grid=both,
        grid style={line width=.4pt, draw=gray!10},
        mark options={solid,scale=1.0},
        major grid style={line width=.2pt,draw=gray!50},
        minor tick num=5,
        xlabel={SNR (dB)},
        ylabel={BLER},
        legend style={at={(0.6, 0.6), font=\footnotesize},anchor=south west},
        mark repeat={2}
      ]
        \addplot[olive, mark=square*] table [x=snr, y=qpsk, col sep=comma] {figs/rbf_sim_bler_vs_snr.csv};
        \addplot[red, mark=+] table [x=snr, y=agrell, col sep=comma] {figs/rbf_sim_bler_vs_snr.csv};
        \addplot[dashed, orange, mark=triangle*] table [x=snr, y=sl-pil, col sep=comma] {figs/rbf_sim_bler_vs_snr.csv};
        \addplot[dotted, orange, mark=oplus*] table [x=snr, y=sl-rtn, col sep=comma] {figs/rbf_sim_bler_vs_snr.csv};
        \addplot[dashed, blue, mark=diamond*] table [x=snr, y=rl-pil, col sep=comma] {figs/rbf_sim_bler_vs_snr.csv};
        \addplot[dotted, blue, mark=pentagon*] table [x=snr, y=rl-rtn, col sep=comma] {figs/rbf_sim_bler_vs_snr.csv};

      \addlegendentry{QPSK}
      \addlegendentry{Agrell16}
      \addlegendentry{Model-aware - Equalized}
      \addlegendentry{Model-aware - Not Equalized}
      \addlegendentry{Model-free - Equalized}
      \addlegendentry{Model-free - Not Equalized}

        \end{axis}

        \end{tikzpicture}
        \caption{\gls{RBF} channel}
        \label{fig:rbf_err_vs_snr}
    \end{subfigure}

    \caption{BLER of the approaches over \gls{AWGN} and \gls{RBF} channels}
    \label{fig:err_vs_snr}
\end{figure*}

Fig.~\ref{fig:err_vs_snr} shows the \gls{BLER} of the compared schemes over the \gls{AWGN} and \gls{RBF} channels.
These results demonstrate that training using the model-free algorithm achieves a similar \gls{BLER} as with the model-aware scheme.
For the \gls{AWGN} channel, the model-aware and alternating algorithms outperform \gls{QPSK}, but not Agrell. This is not surprising as Agrell is a highly optimized solution of the sphere packing problem, leading to probably optimal performance.
However, the learning-based algorithms outperform both \gls{QPSK} and Agrell on the \gls{RBF} channel, showing their ability to learn robust schemes \gls{wrt} channel estimation errors.

Fig.~\ref{fig:rbf_err_vs_snr} also reveals that equalizing the received symbols using a pilot before feeding them to the receiver \gls{NN} leads to the same performance as without prior equalization, but with the additional transformer network.
Interestingly, for \glspl{SNR} above the training SNR, a ``saturation effect'' appears in Fig.~\ref{fig:rbf_err_vs_snr} for both model-aware and model-free methods.

\begin{figure*}
\centering
    \begin{subfigure}{0.45\linewidth}
        \begin{tikzpicture}
            \begin{axis}[%
            name=plot1,
            grid=both,
            grid style={line width=.6pt, draw=gray!10},
            only marks,
            width=0.6\textwidth,
            height=0.6\textwidth,
            xmin=-5, xmax=5,
            ymin=-5, ymax=5,
            xtick={-4,0,4},
            ytick={-4,0,4},
            xticklabels={$-4$, $0$, $4$},
            yticklabels={$-4$, $0$, $4$},
            extra x ticks={-2, 2},
            extra y ticks={-2, 2},
            extra x tick labels={},
            extra y tick labels={}
            ]
                \addplot[
                        scatter,
                        only marks,
                        scatter/classes={ 
                            0={mark=square,green},
                            1={mark=triangle,black},
                            2={mark=o,blue},
                            3={mark=x,red},
                            4={mark=diamond,yellow},
                            5={mark=pentagon,magenta},
                            6={mark=star,orange},
                            7={mark=otimes,brown}
                        },
                    scatter src=explicit symbolic
                ] table [x=x1, y=y1, col sep=comma, meta index=0] {figs/X_CONST.csv};
            \end{axis}
            \begin{axis}[%
            name=plot2,
            grid=both,
            grid style={line width=.6pt, draw=gray!10},
            width=0.6\textwidth,
            height=0.6\textwidth,
            only marks,
            at=(plot1.right of south east), anchor=left of south west,
            xmin=-5, xmax=5,
            ymin=-5, ymax=5,
            xtick={-4,0,4},
            ytick={-4,0,4},
            xticklabels={$-4$, $0$, $4$},
            yticklabels={$-4$, $0$, $4$},
            extra x ticks={-2, 2},
            extra y ticks={-2, 2},
            extra x tick labels={},
            extra y tick labels={}
            ]
                \addplot[
                        scatter,
                        only marks,
                        scatter/classes={ 
                            0={mark=square,green},
                            1={mark=triangle,black},
                            2={mark=o,blue},
                            3={mark=x,red},
                            4={mark=diamond,yellow},
                            5={mark=pentagon,magenta},
                            6={mark=star,orange},
                            7={mark=otimes,brown}
                        },
                    scatter src=explicit symbolic
                ] table [x=x2, y=y2, col sep=comma, meta index=0] {figs/X_CONST.csv};
            \end{axis}
        \end{tikzpicture}
        \caption{Transmitted signal}
        \label{fig:const_x}
    \end{subfigure}
    \begin{subfigure}{0.45\linewidth}
        \begin{tikzpicture}
            \begin{axis}[%
            name=plot1,
            grid=both,
            grid style={line width=.6pt, draw=gray!10},
            only marks,
            width=0.6\textwidth,
            height=0.6\textwidth,
            xmin=-5, xmax=5,
            ymin=-5, ymax=5,
            xtick={-4,0,4},
            ytick={-4,0,4},
            xticklabels={$-4$, $0$, $4$},
            yticklabels={$-4$, $0$, $4$},
            extra x ticks={-2, 2},
            extra y ticks={-2, 2},
            extra x tick labels={},
            extra y tick labels={}
            ]
                \addplot[
                    scatter,
                    only marks,
                    scatter/classes={ 
                        0={mark=square,green},
                        1={mark=triangle,black},
                        2={mark=o,blue},
                        3={mark=x,red},
                        4={mark=diamond,yellow},
                        5={mark=pentagon,magenta},
                        6={mark=star,orange},
                        7={mark=otimes,brown}
                    },
                    scatter src=explicit symbolic
                ] table [x=x1, y=y1, col sep=comma, meta index=0] {figs/Y_CONST.csv};
            \end{axis}
            \begin{axis}[%
            name=plot2,
            grid=both,
            grid style={line width=.6pt, draw=gray!10},
            width=0.6\textwidth,
            height=0.6\textwidth,
            only marks,
            at=(plot1.right of south east), anchor=left of south west,
            xmin=-5, xmax=5,
            ymin=-5, ymax=5,
            xtick={-4,0,4},
            ytick={-4,0,4},
            xticklabels={$-4$, $0$, $4$},
            yticklabels={$-4$, $0$, $4$},
            extra x ticks={-2, 2},
            extra y ticks={-2, 2},
            extra x tick labels={},
            extra y tick labels={}
            ]
                \addplot[
                    scatter,
                    only marks,
                    scatter/classes={ 
                        0={mark=square,green},
                        1={mark=triangle,black},
                        2={mark=o,blue},
                        3={mark=x,red},
                        4={mark=diamond,yellow},
                        5={mark=pentagon,magenta},
                        6={mark=star,orange},
                        7={mark=otimes,brown}
                    },
                    scatter src=explicit symbolic
                ] table [x=x2, y=y2, col sep=comma, meta index=0] {figs/Y_CONST.csv};
            \end{axis}
        \end{tikzpicture}
        \caption{Received signal}
        \label{fig:const_y}
    \end{subfigure}
    \vspace{0.5\baselineskip}

    \begin{subfigure}{0.45\linewidth}
        \begin{tikzpicture}
            \begin{axis}[%
            name=plot1,
            grid=both,
            grid style={line width=.6pt, draw=gray!10},
            only marks,
            width=0.6\textwidth,
            height=0.6\textwidth,
            xmin=-5, xmax=5,
            ymin=-5, ymax=5,
            xtick={-4,0,4},
            ytick={-4,0,4},
            xticklabels={$-4$, $0$, $4$},
            yticklabels={$-4$, $0$, $4$},
            extra x ticks={-2, 2},
            extra y ticks={-2, 2},
            extra x tick labels={},
            extra y tick labels={}
            ]
                \addplot[
                    scatter,
                    only marks,
                    scatter/classes={ 
                        0={mark=square,green},
                        1={mark=triangle,black},
                        2={mark=o,blue},
                        3={mark=x,red},
                        4={mark=diamond,yellow},
                        5={mark=pentagon,magenta},
                        6={mark=star,orange},
                        7={mark=otimes,brown}
                    },
                    scatter src=explicit symbolic
                ] table [x=x1, y=y1, col sep=comma, meta index=0] {figs/Y_HAT_CONST.csv};
            \end{axis}
            \begin{axis}[%
            name=plot2,
            grid=both,
            grid style={line width=.6pt, draw=gray!10},
            width=0.6\textwidth,
            height=0.6\textwidth,
            only marks,
            at=(plot1.right of south east), anchor=left of south west,
            xmin=-5, xmax=5,
            ymin=-5, ymax=5,
            xtick={-4,0,4},
            ytick={-4,0,4},
            xticklabels={$-4$, $0$, $4$},
            yticklabels={$-4$, $0$, $4$},
            extra x ticks={-2, 2},
            extra y ticks={-2, 2},
            extra x tick labels={},
            extra y tick labels={}
            ]
                \addplot[
                    scatter,
                    only marks,
                    scatter/classes={ 
                        0={mark=square,green},
                        1={mark=triangle,black},
                        2={mark=o,blue},
                        3={mark=x,red},
                        4={mark=diamond,yellow},
                        5={mark=pentagon,magenta},
                        6={mark=star,orange},
                        7={mark=otimes,brown}
                    },
                    scatter src=explicit symbolic
                ] table [x=x2, y=y2, col sep=comma, meta index=0] {figs/Y_HAT_CONST.csv};
            \end{axis}
        \end{tikzpicture}
        \caption{Transformed signal}
        \label{fig:const_yhat}
    \end{subfigure}
    \begin{subfigure}{0.45\linewidth}
        \begin{tikzpicture}
            \begin{axis}[%
            name=plot1,
            grid=both,
            grid style={line width=.6pt, draw=gray!10},
            width=0.6\textwidth,
            height=0.6\textwidth,
            only marks,
            xmin=-80, xmax=80,
            ymin=-80, ymax=80,
            xtick={-50,0,50},
            ytick={-50,0,50},
            xticklabels={$-5$, $0$, $5$},
            yticklabels={$-5$, $0$, $5$},
            extra x ticks={-75, -25, 25, 75},
            extra y ticks={-75, -25, 25, 75},
            extra x tick labels={},
            extra y tick labels={}
            ]
                \addplot[
                    scatter,
                    only marks,
                    scatter/classes={  0={mark=square,green},
                                       1={mark=triangle,black},
                                       2={mark=o,blue},
                                       3={mark=x,red},
                                       4={mark=diamond,yellow},
                                       5={mark=pentagon,magenta},
                                       6={mark=star,orange},
                                       7={mark=otimes,brown}
                                  },
                    scatter src=explicit symbolic
                ] table [x=x, y=y, col sep=comma, meta index=0] {figs/Y_TSNE.csv};
            \end{axis}
            \begin{axis}[%
            name=plot2,
            grid=both,
            grid style={line width=.6pt, draw=gray!10},
            width=0.6\textwidth,
            height=0.6\textwidth,
            only marks,
            at=(plot1.right of south east), anchor=left of south west,
            xmin=-80, xmax=80,
            ymin=-80, ymax=80,
            xtick={-50,0,50},
            ytick={-50,0,50},
            xticklabels={$-5$, $0$, $5$},
            yticklabels={$-5$, $0$, $5$},
            extra x ticks={-75, -25, 25, 75},
            extra y ticks={-75, -25, 25, 75},
            extra x tick labels={},
            extra y tick labels={}
            ]
                \addplot[
                    scatter,
                    only marks,
                    scatter/classes={  0={mark=square,green},
                                       1={mark=triangle,black},
                                       2={mark=o,blue},
                                       3={mark=x,red},
                                       4={mark=diamond,yellow},
                                       5={mark=pentagon,magenta},
                                       6={mark=star,orange},
                                       7={mark=otimes,brown}
                                  },
                    scatter src=explicit symbolic
                ] table [x=x, y=y, col sep=comma, meta index=0] {figs/Y_HAT_TSNE.csv};
            \end{axis}
        \end{tikzpicture}
        \caption{t-SNE. Left: Received -- Right: Transformed}
        \label{fig:tsne}
    \end{subfigure}
    \caption{Constellations corresponding to the transmitted signal, the received signal, the transformed signal, and t-SNE visualization, with alternating training, \gls{RBF} channel, $M = 8$ and $N = 2$. \rev{In figures~(a),~(b) and~(c), the left and right graphs correspond to the first and second channel use, respectively.}}
    \label{fig:const}
\end{figure*}
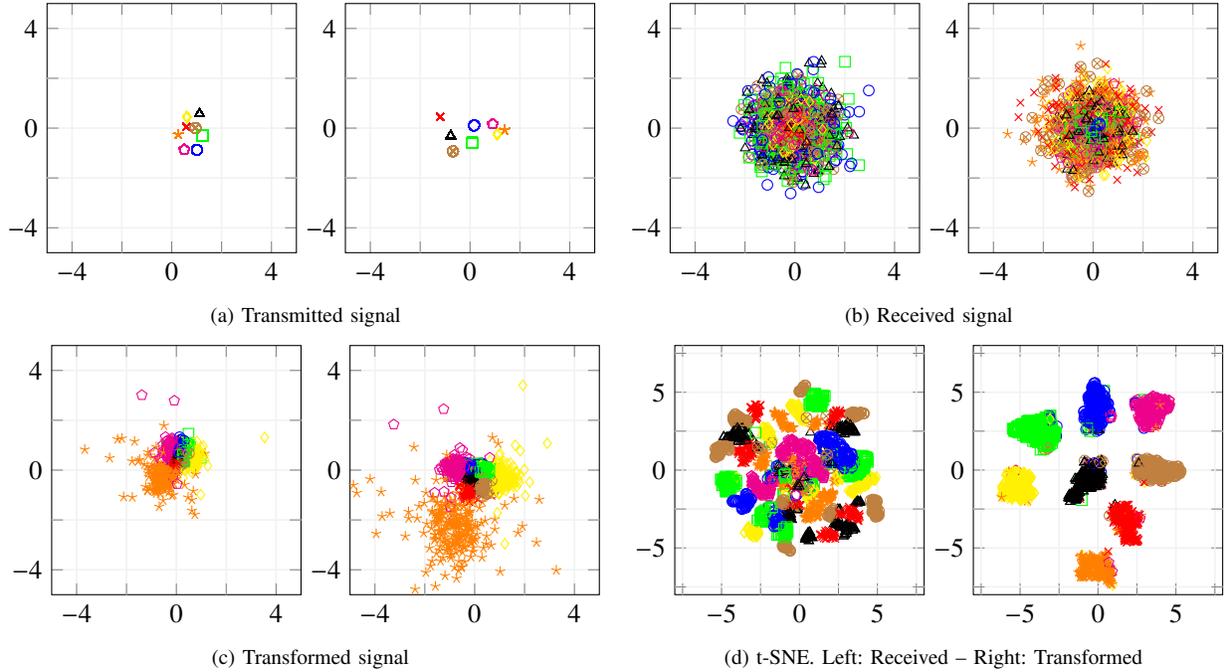

To get insight into how the receiver achieves signal detection for the \gls{RBF} channel when no prior equalization is done, the t-SNE~\cite{tsne} visualization algorithm was used.
t-SNE is a probabilistic dimensionality reduction algorithm which maps high-dimensional points into low dimensions (typically 2 or 3) to enable visualization.
Moreover, objects close to each other in the high-dimensional space are mapped to points close to each other in the low-dimensional representation with high probability.
For readability, we used $M=8$ and $N=2$ for this study, and consider only alternating training.
Fig.~\ref{fig:const} shows the constellation corresponding to the transmitted, received, and transformed signals, as well as the t-SNE visualization.
It can be seen in Fig.~\ref{fig:const_x} that the transmitted constellation of the first channel use is not centered, which indicates that the transmitter has learned to add super-imposed pilots which allow the receiver to detect the sent message.
The same observation was made in~\cite{dorner2017deep}.
The received signal (in Fig.~\ref{fig:const_y}) does not seem to present any structure from which it is easy to extract information.
Visualization of this signal using t-SNE also fails to reveals clusters, as shown in Fig.~\ref{fig:tsne} (left).
On the other hand, t-SNE reveals clusters corresponding to the 8 possible messages when considering the transformed signal, as shown in Fig.~\ref{fig:tsne} (right).
One can hence speculate that the transformer network uses the super-imposed pilots to transform the received signal such that signals corresponding to different messages are separable by the discriminative network.
Interestingly, visual inspection of the constellations associated to the transformed signal, shown in Fig.~\ref{fig:const_yhat}, does not reveal any obvious separations between the messages, meaning that clusters appear only in high dimensions.
A key point to understand the saturation effect in Fig.~\ref{fig:rbf_err_vs_snr} is that, as opposite to when explicit prior estimation/equalization is performed, the receiver does not have a perfect knowledge of the pilot signal, leading to a permanent residual error.
This residual error is negligible at \glspl{SNR} below the training \gls{SNR}, but becomes significant at \glspl{SNR} considerably higher, leading to the observed saturation effect.

\subsection{Evaluation of the convergence rate}

\rev{
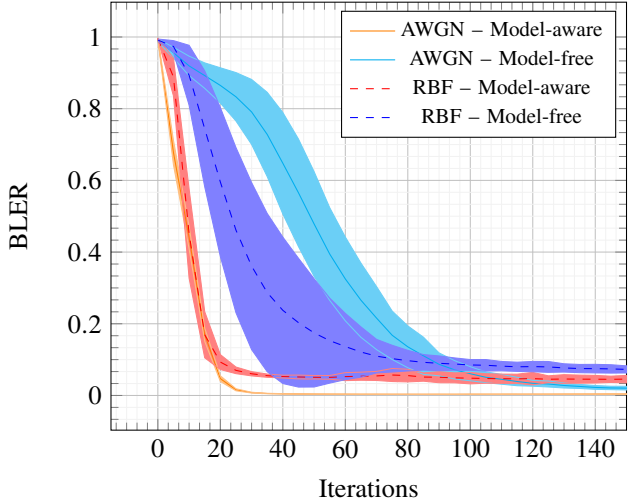
\begin{figure}
    \centering
\begin{tikzpicture}
  \begin{axis}[
    grid=both,
    grid style={line width=.1pt, draw=gray!10},
    major grid style={line width=.2pt,draw=gray!50},
    minor tick num=5,
    xlabel={Iterations},
    ylabel={BLER},
    legend style={font=\footnotesize},
    xmax=150
]
    \addplot[orange] table [x=iterations, y=awgn_sl, col sep=comma] {figs/training_2.csv}; \addlegendentry{AWGN -- Model-aware}
    \addplot[orange!50, name path=awgn_sl_up,forget plot] table [x=iterations, y expr=(\thisrowno{1} + \thisrowno{3}), col sep=comma] {figs/training_2.csv};
    \addplot[orange!50, name path=awgn_sl_down,forget plot] table [x=iterations, y expr=(\thisrowno{1} - \thisrowno{3}), col sep=comma] {figs/training_2.csv};
    \addplot[orange!50,forget plot] fill between[of=awgn_sl_up and awgn_sl_down];
    \addplot[cyan] table [x=iterations, y=awgn_rl, col sep=comma] {figs/training_2.csv}; \addlegendentry{AWGN -- Model-free}
    \addplot[cyan!50, name path=awgn_rl_up,forget plot] table [x=iterations, y expr=(\thisrowno{2} + \thisrowno{4}), col sep=comma] {figs/training_2.csv};
    \addplot[cyan!50, name path=awgn_rl_down,forget plot] table [x=iterations, y expr=(\thisrowno{2} - \thisrowno{4}), col sep=comma] {figs/training_2.csv};
    \addplot[cyan!50,forget plot] fill between[of=awgn_rl_up and awgn_rl_down];
    \addplot[red, dashed] table [x=iterations, y=rbf_sl, col sep=comma] {figs/training_2.csv}; \addlegendentry{RBF -- Model-aware}
    \addplot[red!50, name path=rbf_sl_up,forget plot] table [x=iterations, y expr=(\thisrowno{5} + \thisrowno{7}), col sep=comma] {figs/training_2.csv};
    \addplot[red!50, name path=rbf_sl_down,forget plot] table [x=iterations, y expr=(\thisrowno{5} - \thisrowno{7}), col sep=comma] {figs/training_2.csv};
    \addplot[red!50,forget plot] fill between[of=rbf_sl_up and rbf_sl_down];
    \addplot[blue, dashed] table [x=iterations, y=rbf_rl, col sep=comma] {figs/training_2.csv}; \addlegendentry{RBF -- Model-free}
    \addplot[blue!50, name path=rbf_rl_up,forget plot] table [x=iterations, y expr=(\thisrowno{6} + \thisrowno{8}), col sep=comma] {figs/training_2.csv};
    \addplot[blue!50, name path=rbf_rl_down,forget plot] table [x=iterations, y expr=(\thisrowno{6} - \thisrowno{8}), col sep=comma] {figs/training_2.csv};
    \addplot[blue!50,forget plot] fill between[of=rbf_rl_up and rbf_rl_down];
\end{axis}
\end{tikzpicture}
    \caption{Evolution of the BLER during the 150 first training iterations}
    \label{fig:training}
\end{figure}
}

\rev{
The evolution of the \gls{BLER} of the model-aware and the model-free algorithm is shown in Fig.~\ref{fig:training}, averaged over 200 seeds used for initialization of the \gls{NN} weights.
Because each iteration of model-free training consists of ten gradient descent steps on the receiver and the transmitter, respectively, we assume for fairness that each iteration of model-aware training corresponds to ten gradient descent steps on the end-to-end system.
As significant differences in convergence speed are only observed at the beginning of the training process, only the first 150 iterations are shown for readability.
Shaded areas around the curves correspond to one standard deviation in each direction. For the \gls{RBF} channel, only the case with no prior equalization is shown.
For both the \gls{AWGN} and \gls{RBF} channels, the model-aware method leads to faster convergence compared to the alternating method.
This is expected since the true loss gradient is used to train the transmitter, in contrast to an approximation of the loss gradient when no model of the channel is provided.
Significant differences in convergence speed are only observed at the beginning of the training process, showing that model-free training does not lead to a significantly slower convergence compared to model-aware training.
Moreover, as previous results show, once properly trained, both model-free and model-aware training achieve the same performance.
}

\subsection{Evaluation with a noisy feedback link}

The proposed algorithm requires a feedback link to send the per-example losses computed by the receiver to the transmitter (line~\ref{lst:te_loss_send_s} in Algorithm~\ref{alg:training_tx}).
Knowledge of these losses is required at the transmitter for training.
However, in practice, this feedback link may be error-prone, or the losses may be quantized.
Both situations result in erroneous losses used for gradient estimation.
Their impact will be investigated in the remainder of this section.

\begin{figure}
    \centering
    \begin{tikzpicture}
      \begin{axis}[
        grid=both,
        grid style={line width=.4pt, draw=gray!10},
        mark options={solid,scale=1.2},
        major grid style={line width=.2pt,draw=gray!50},
        minor tick num=5,
        xlabel={SNR\textsubscript{fb} (dB)},
        ylabel={BLER},
        xmin=-6,
        xmax=11
      ]
      \addplot [dashed, black, mark=none] coordinates {
            (-7, 0.0006489753723144531)
            (12, 0.0006489753723144531)
        };
        \addplot[blue, mark=diamond*] table [x=snr, y=noisy, col sep=comma] {figs/RL_AUTOENC_NOISE_FB.csv};

        \addlegendentry{Noiseless feedback}
        \addlegendentry{Noisy feedback}

        \end{axis}

        \end{tikzpicture}

    \caption{BLER when training with a noisy feedback link}
    \label{fig:noisy_fb}
\end{figure}
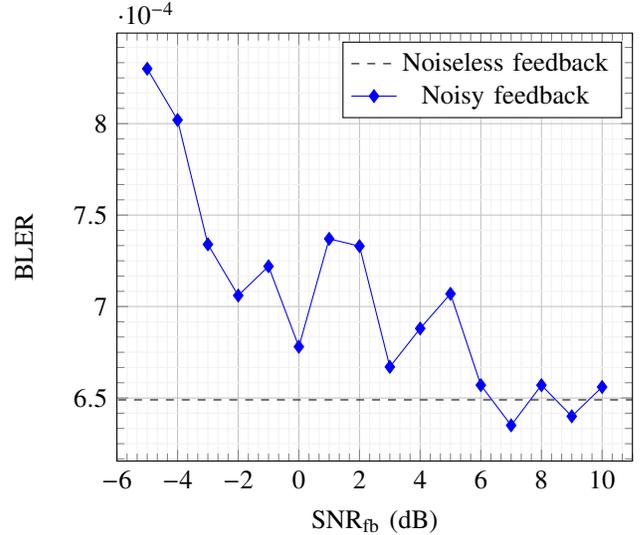

For simplicity, errors are modeled by adding independent zero-mean Gaussian noise with variance $\sigma_e^2$ to the losses, i.e., $\tilde{l} = l + e$, where $\tilde{l}$ denotes an erroneous loss, and $e \thicksim \Nc(0, \sigma_e^2)$.
The feedback link \gls{SNR} is defined as
\begin{equation}
    \text{SNR\textsubscript{fb}} = \frac{\EE \LP l^2 \RP}{\sigma_e^2}.
\end{equation}
Fig.~\ref{fig:noisy_fb} shows the \gls{BLER} of the communication system trained using the alternating method with erroneous feedback and for various values of \gls{SNR}\textsubscript{fb}.
The number of messages $M$ was set to 256, the number of channel uses $N$ to 4, and both training and evaluation were performed at an SNR of $10\:$dB over an \gls{AWGN} channel.
Surprisingly, erroneous losses have a negligible impact on the \gls{BLER} for positive values of \gls{SNR}\textsubscript{fb}.
Moreover, for \gls{SNR}\textsubscript{fb} higher than $6\:$dB, no \gls{BLER} increase is observed.
This result suggests that the proposed training algorithm is robust to erroneous feedback, which is encouraging with regards to its practical use.

\subsection{Joint source-channel coding of images}

By modifying transmitter and receiver \gls{NN} architectures in Fig.~\ref{fig:nn_gen_arch} to real-valued vector inputs, one can train communication systems for other tasks than transmitting messages drawn from a discrete set.
One recent example is joint source-channel coding for images, which was recently addressed in~\cite{bourtsoulatze2018deep,Choi}.
We will now show that model-free training can be applied in these settings, too.

\begin{figure}
    \centering
  \begin{subfigure}{0.9\linewidth}
    \centering
    \includegraphics[width=\linewidth]{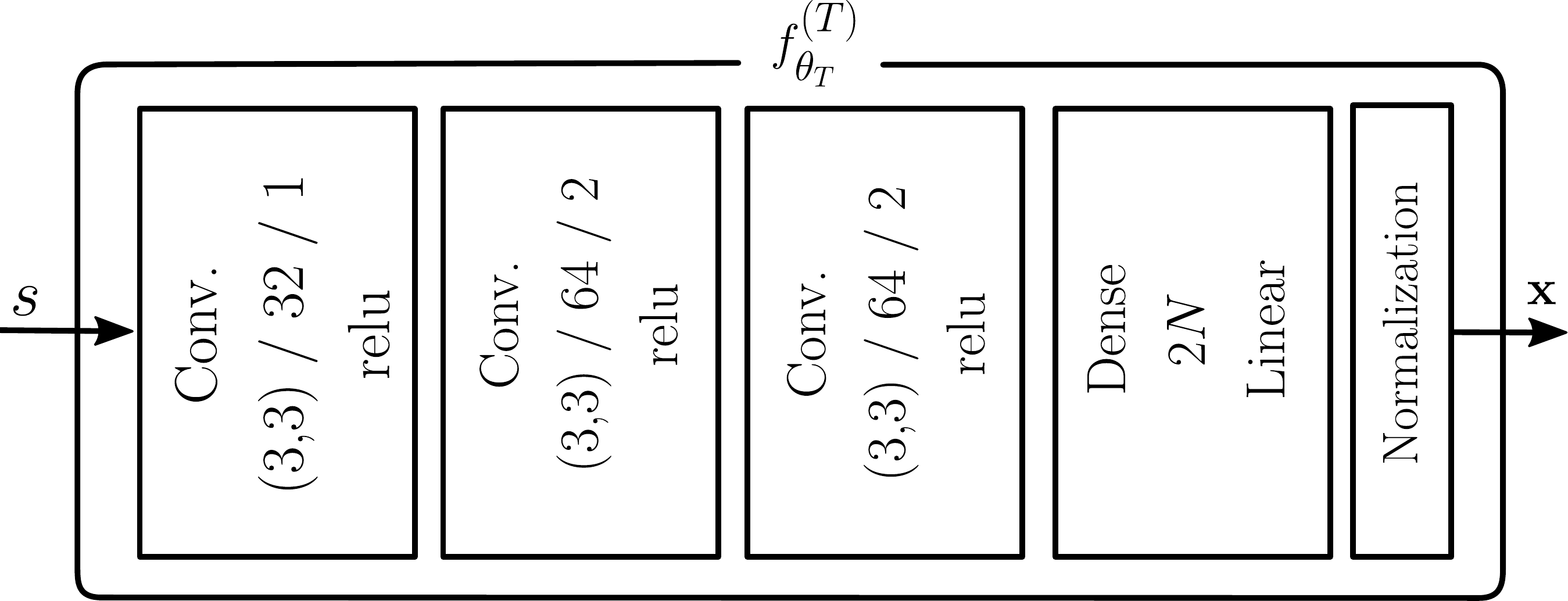}
    \subcaption{Transmitter architecture}
    \label{fig:mnist_tx}
  \end{subfigure}

  \begin{subfigure}{0.8\linewidth}
    \centering
    \includegraphics[width=\linewidth]{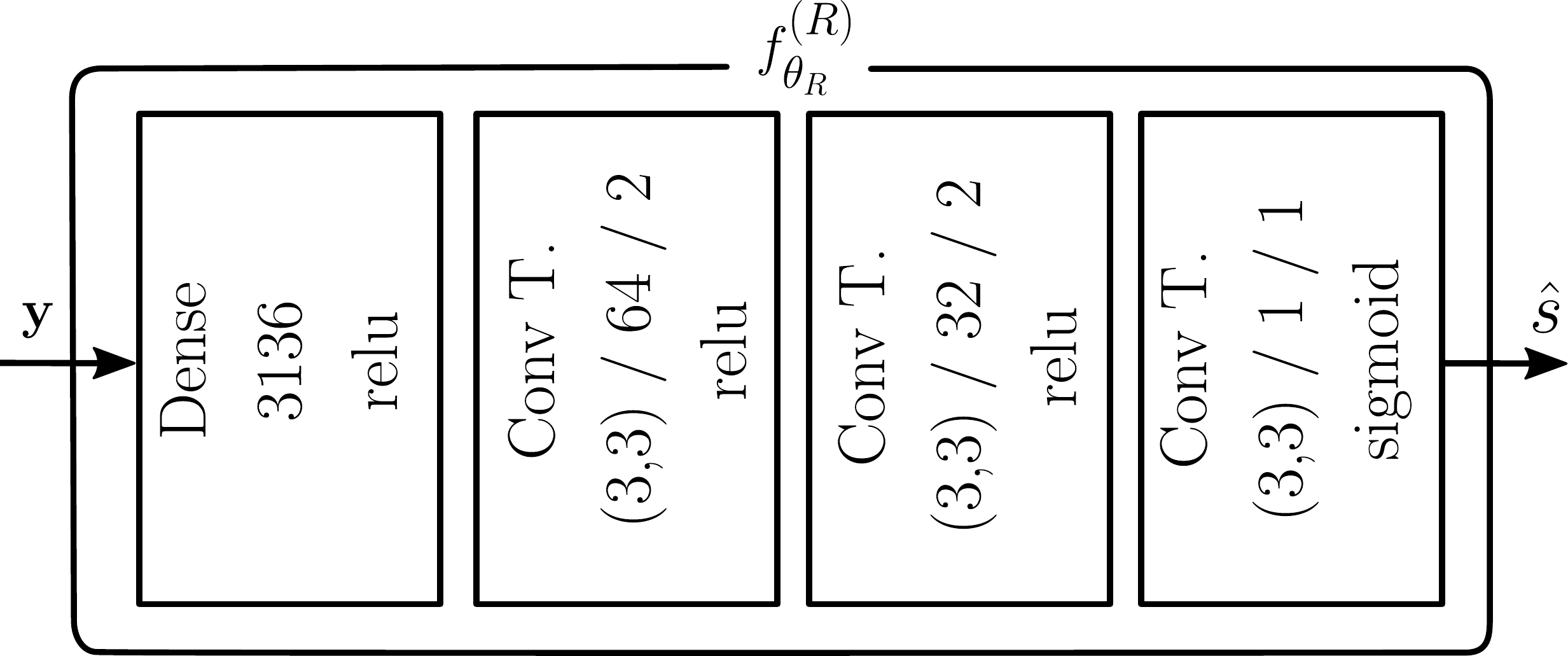}
    \subcaption{Receiver architecture}
    \label{fig:mnist_rx}
  \end{subfigure}
  \caption{Transmitter and receiver architectures used for transmission of images. In convolutional or transposed convolutional layers, parameters are indicated as ``kernel size/filters/stride''.}
  \label{fig:mnist_ae}
\end{figure}

The MNIST~\cite{mnist} dataset of handwritten digits is considered, in which each example is a 28-by-28 grayscale image.
The generic architectures of transmitter and receiver presented in Section~\ref{sec:gen_arch} are adapted to image processing.
Especially, the receiver does not output a probability vector in this application, but an image with the same dimensions as the input.
The used autoencoder architecture is shown in Fig.~\ref{fig:mnist_ae}.
Convolutional layers~\cite{Goodfellow-et-al-2016-Book} are used in the transmitter to exploit the spatial structure of images, with strides higher than one to achieve compression.
A dense layer is leveraged to generate the channel symbols, which are then normalized to ensure the power constraint.
The receiver is designed symmetrical to the transmitter, and leverages transposed convolutional layers to reconstruct images from low-dimensional inputs.
Sigmoid activation functions are used in the last convolutional layer to output grayscale pixel values in the range $[0,1]$.
The loss function is the \gls{MSE}, and reconstruction quality is measured by the \gls{PSNR}, defined as:
\begin{equation}
    \text{PSNR} = \frac{1}{\text{MSE}}.
\end{equation}

An \gls{AWGN} channel is considered with $N = 10$ channel uses and training is performed at an \gls{SNR} of $10\:$dB.
We observed experimentally that training using the model-free algorithm requires larger batch sizes for the transmitter than for the receiver.
We believe that this unwanted effect is due to an increased estimator variance as the number of transmitter outputs increases.
\rev{Similar effects were observed in \gls{RL}, where the difficulty of training systems with continuous and high dimensional output spaces is well-known in~\cite{lilli16}.}
Note that this effect is different from the one described for \gls{SPSA} (see Appendix~\ref{app:spsa}), for which the estimator variance increases with the number of trainable parameters.
Indeed, the number of trainable parameters typically increases at a faster rate than the number of outputs.
As an example, the transmitter architecture used in this study has only twenty outputs, but more than ten thousand trainable parameters.
This effect can however be compensated for by larger batch sizes.
\begin{figure}
    \centering
    \begin{tikzpicture}
        \begin{axis}[
            grid=both,
            grid style={line width=.4pt, draw=gray!10},
            mark options={solid,scale=1.2},
            major grid style={line width=.2pt,draw=gray!50},
            minor tick num=5,
            xlabel={SNR (dB)},
            ylabel={PSNR (dB)},
            legend style={at={(0.5, 0.2)},anchor=south west},
            mark repeat={2}
        ]
            \addplot[dashed, blue, mark=triangle*] table [x=snr, y=sl_128, col sep=comma] {figs/mnist_autoenc_awgn.csv};
            \addplot[dashed, orange, mark=diamond*] table [x=snr, y=rl_128, col sep=comma] {figs/mnist_autoenc_awgn.csv};

            \addlegendentry{Model-aware}
            \addlegendentry{Model-free}

        \end{axis}

    \end{tikzpicture}

    \caption{PSNR for MNIST reconstruction over an \gls{AWGN} channel}
    \label{fig:mnist_psnr}
\end{figure}
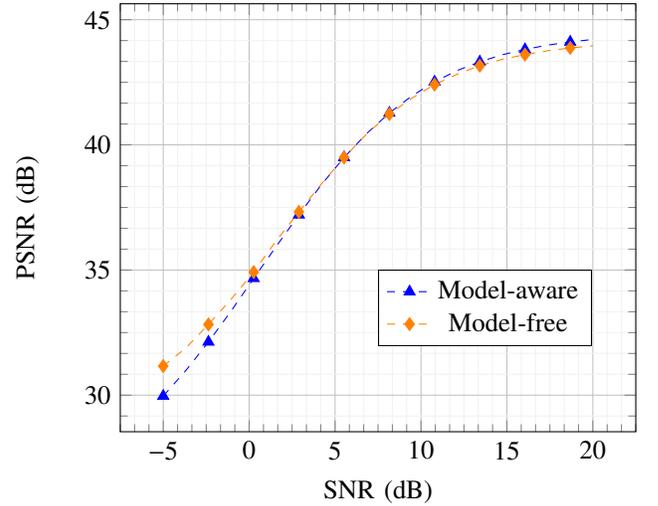
Therefore, using the alternating approach, training of the transmitter was done with a fixed batch size of 128 examples, while training of the receiver was done with a progressively increasing batch size, up to 128.
When training with channel model knowledge, the batch size was progressively increased up to 128.
Fig.~\ref{fig:mnist_psnr} shows the \gls{PSNR} of the model-free and model-aware algorithms as a function of the \gls{SNR}.
It can be seen that model-free training incurs no performance penalty, showing that alternating training can handle complex transmitter architectures and end-to-end tasks.
\begin{figure}
    \centering
    \begin{subfigure}{.25\linewidth}
    \centering
        \begin{tabular}{c}
            \footnotesize Original
        \end{tabular}
    \end{subfigure}
    \begin{subfigure}{0.20\linewidth}
        \centering
        \includegraphics[width=\linewidth]{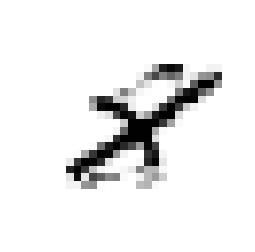}
    \end{subfigure}
    \begin{subfigure}{0.20\linewidth}
        \centering
        \includegraphics[width=\linewidth]{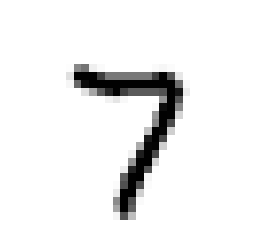}
    \end{subfigure}
    \begin{subfigure}{0.20\linewidth}
        \centering
        \includegraphics[width=\linewidth]{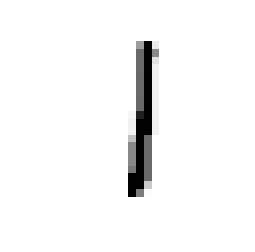}
    \end{subfigure}

    \begin{subfigure}{.25\linewidth}
    \centering
        \begin{tabular}{c}
            \footnotesize Reconsc.
        \end{tabular}
    \end{subfigure}
    \begin{subfigure}{0.20\linewidth}
        \centering
        \includegraphics[width=\linewidth]{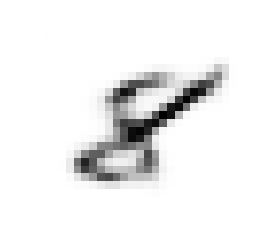}
        \caption{$28.5\:$dB}
    \end{subfigure}
    \begin{subfigure}{0.20\linewidth}
        \centering
        \includegraphics[width=\linewidth]{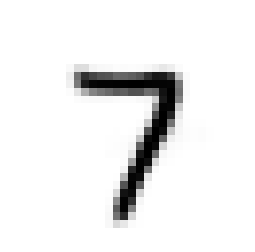}
        \caption{$45.4\:$dB}
    \end{subfigure}
    \begin{subfigure}{0.20\linewidth}
        \centering
        \includegraphics[width=\linewidth]{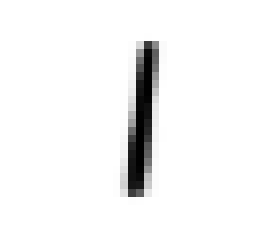}
        \caption{$72.6\:$dB}
    \end{subfigure}

  \caption{Examples of MNIST image reconstructions over an \gls{AWGN} channel with an \gls{SNR} of $10\:$dB. The captions indicate the corresponding \gls{PSNR}, with increasing reconstruction quality.}
  \label{fig:mnist_ex}
\end{figure}
Examples of images generated by the autoencoder trained with the alternating algorithm are shown in Fig.~\ref{fig:mnist_ex}.

\subsection{Evaluation on a fiber-optical channel}

\rev{
\begin{figure}
    \centering
    \begin{tikzpicture}
      \begin{axis}[
        ymode=log,
        grid=both,
        grid style={line width=.4pt, draw=gray!10},
        mark options={solid,scale=1.2},
        major grid style={line width=.2pt,draw=gray!50},
        minor tick num=5,
        xlabel={SNR (dB)},
        ylabel={SER},
        legend style={at={(0.1, 0.2)},anchor=south west}
      ]
        \addplot[dashed, orange, mark=triangle*] table [x=snr, y=supervised, col sep=comma] {figs/optics.csv};
        \addplot[dashed, blue, mark=diamond*] table [x=snr, y=alternating, col sep=comma] {figs/optics.csv};
      \addlegendentry{Model-aware}
      \addlegendentry{Model-free}
        \end{axis}
        \end{tikzpicture}
    \caption{SER of the model-aware and model-free algorithms on a fiber-optical channel}
    \label{fig:optics}
\end{figure}
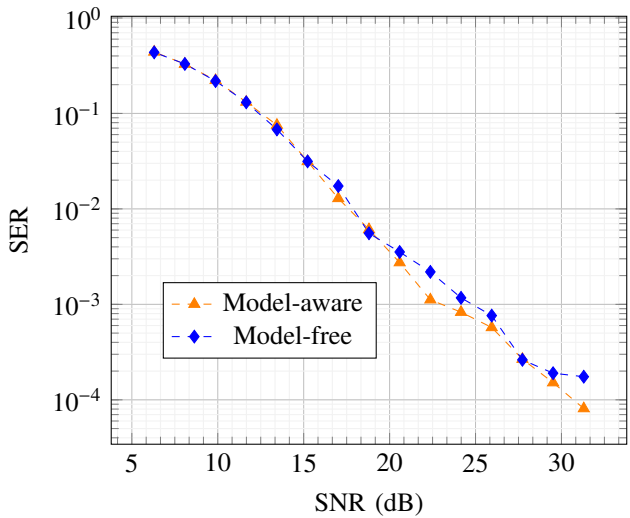
}

We conclude this section by evaluating the alternating algorithm on a simplified fiber-optical channel model. The simplified model is obtained from the nonlinear Schr\"{o}dinger equation, by neglecting dispersion~\cite{li2018fiberautoenc}, which leads to a recursive per-sample model
\begin{equation}
    \xv_k =
    \begin{cases}
        \xv_{k-1} \exp{j \frac{L \gamma \abs{\xv_{k-1}}^2}{K} } + \nv_k & \text{for } 1 \leq k < K\\
        \xv                                             & \text{for } k = 0
    \end{cases}
\end{equation}
where $\xv \in \CC^{N}$ is the channel input, and $\yv = \xv_K$ the channel output.
The complex-valued channel input was created by interpreting the first $N$ elements of the transmitter \gls{NN} output as the real and the second part as the imaginary part of $\xv$.
The inverse operation was performed at the receiver side.
$L$ is the fiber length, $\gamma$ the nonlinearity parameter, and $\nv_k \thicksim \Cc\Nc(\vec{0}, \frac{\sigma_n^2}{K})$, where $\sigma_n^2$ is the noise power. $K$ is assumed large.

While the generic architectures presented in Section~\ref{sec:gen_arch} are valid for any channel, the architecture introduced in Section~\ref{sec:archi} is not suited to fiber-optical channels.
Instead, the transmitter is made of two hidden dense layers of 64 neurons each with ReLu activation, followed by a hidden dense layer of $2N$ neurons with linear activation.
The normalization layer is such that the average energy per symbol is set to a controllable parameter $P_{in}$.
The receiver is made of 2 hidden dense layers of 64 neurons each with ReLu activation, followed by the output softmax layer.
As in~\cite{li2018fiberautoenc}, the channel parameters were $L = 5000\:$km, $\gamma = 1.27\:$W/km, and $K$ was set to 50, which is sufficient to approximate the asymptotic channel model.
$M$ was set to 16, $N$ to 1, and $\sigma$ to 0.05 for alternating training.
For these parameters, Li et al.~\cite{li2018fiberautoenc} showed that an autoencoder-based communication system trained with model knowledge outperforms 16-QAM modulation under maximum likelihood detection.
Therefore, we take the autoencoder trained with model knowledge as a baseline to evaluate the performance of the alternating algorithm.

Fig.~\ref{fig:optics} shows the \gls{SER} of the autoencoder trained with the model-aware and model-free approaches.
As in~\cite{li2018fiberautoenc}, a separate \gls{NN} is trained for each \gls{SNR} value as the trained \glspl{NN} struggle to generalize to other \gls{SNR} values.
It can be seen that the model-aware and model-free approaches achieve the same \gls{SER}.
These results are encouraging as they illustrate the universality of the alternating training algorithm, which achieves competitive results on both wireless and fiber-optical channels.

\section{Over-the-air experiments}

The first proof-of-concept of an autoencoder-based communication system was described in~\cite{dorner2017deep}.
This prototype was trained using a channel model, and then deployed and evaluated over the actual channel.
One of the main difficulties encountered by the authors of this early work was the difference between the actual channel and the channel model used for training, which led to significant performance loss after deployment.
Most of this mismatch comes from imperfections of the transmitter and receiver hardware, which are hard to model.
The alternating algorithm proposed in this work removes the need for an accurate model of the channel.
In this section, we describe a proof-of-concept using the alternating algorithm to train an autoencoder for communications over wireless and coaxial cable channels.
The performance is compared to that of \gls{QPSK} and Agrell~\cite{Agrell16}.
To the best of our knowledge, this is the first prototype of an autoencoder-based communication system trained over actual channels.

\subsection{Experimental setup}

\begin{figure}
    \centering
      \includegraphics[width=0.9\linewidth]{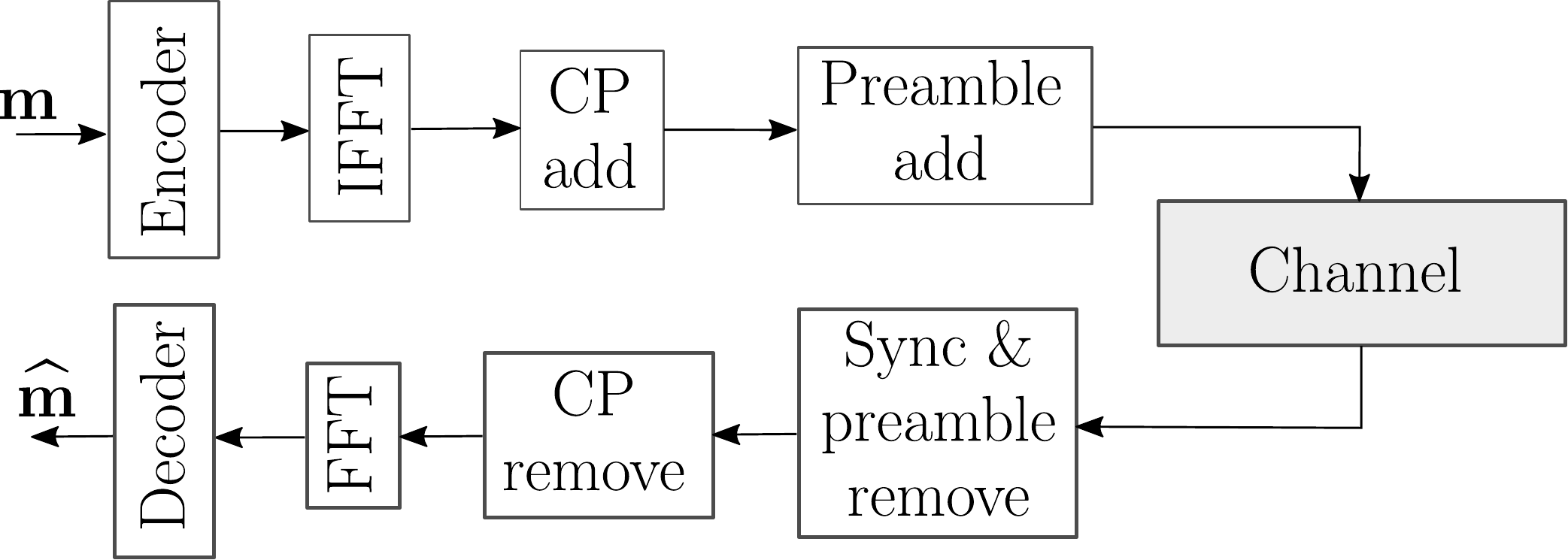}
    \caption{Autoencoder combined with OFDM}
\label{fig:xp_chain}
\end{figure}

Similarly to~\cite{felix2018ofdm}, the autoencoder is extended to \gls{OFDM} with \gls{CP}, as shown in Fig.~\ref{fig:xp_chain}.
A (discrete) \gls{IFFT} of size $w_{FFT}$ is performed on the IQ-symbols generated by the encoder from a set of independent messages $\mv$.
The DC subcarrier, and $G$ guard subcarriers on each side, are not used, leading to $w_{\text{FFT}} - 2G - 1$ data-carrying subcarriers.
To avoid \gls{ISI}, a \gls{CP} of length $l_{\text{CP}}$ is further added.
An additional preamble made of repetitions of the Barker sequence of length 13 is added for synchronization.
On the receiver side, synchronization is performed through correlation with peak detection.
Finally, after removing the \gls{CP}, the input to the decoder is recovered through \gls{FFT}.
The use of \gls{OFDM} with cyclic-prefix enables single-tap equalization.
Similarly to previous evaluations, the autoencoder-based communication system relies on the transmitter and receiver architectures introduced in Section~\ref{sec:archi}, including the transformer network.
\rev{Each message was assigned to a subcarrier, and a single decoder, which includes the transformer network, operates on a per-subcarrier basis.}
It is possible to learn synchronization with additional \glspl{NN}, as done in~\cite{dorner2017deep}.
However, we could not observe any gain for our setup and, therefore, resorted to classical approaches.

\begin{figure}
    \centering
      \includegraphics[width=0.7\linewidth]{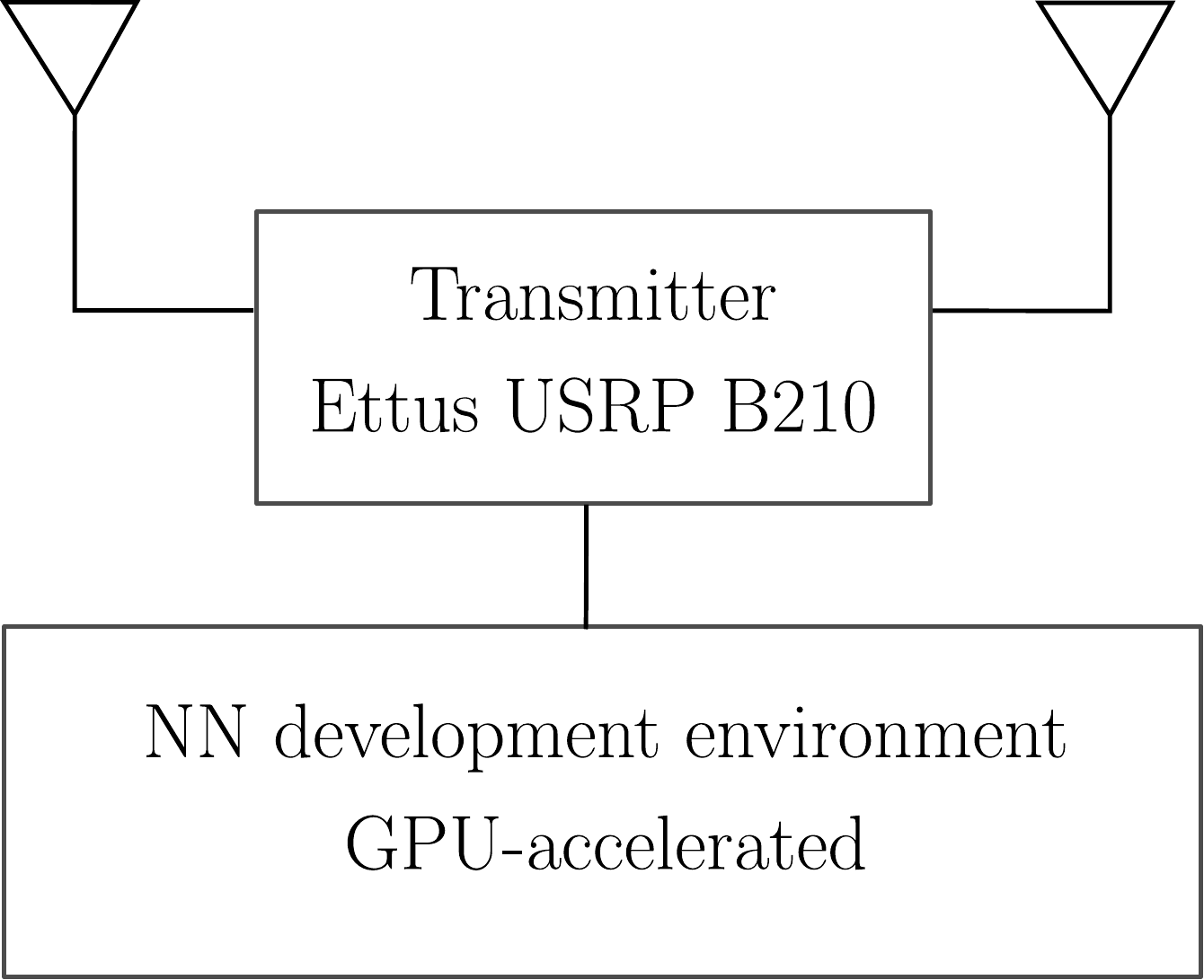}
    \caption{Testbed overview}
\label{fig:testbed_ov}
\end{figure}

Fig.~\ref{fig:testbed_ov} gives an overview of the experimental testbed.
A single Ettus USRP B210 board, connected to a computer equipped with an NVIDIA 1080 consumer class \gls{GPU} and running the TensorFlow framework, was used as transmitter and receiver.
The testbed was deployed in our offices, the antennas had an unobstructed \gls{LOS} path, and were kept unmoved during data transmission.
The carrier frequency was $2.3\:$GHz, and the bandwidth was $5\:$MHz.
The number of subcarriers was $w_{\text{FFT}} = 128$, and the \gls{CP} length $l_{\text{CP}} = 16\:$symbols.
$G = 6$ guard subcarriers were used on each side.
The considered schemes were the same as for the \gls{RBF} channel in Table~\ref{tab:comp_schemes}.

\subsection{Results}

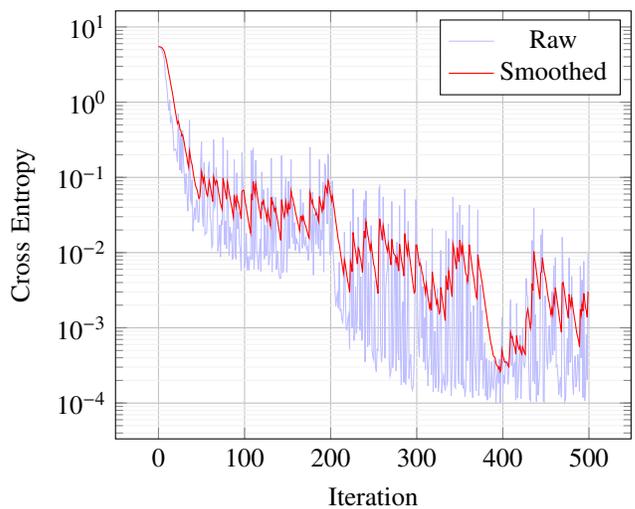
\begin{figure}
	\centering
	\begin{tikzpicture}
	  \begin{axis}[
	    ymode=log,
	    grid=both,
	    grid style={line width=.4pt, draw=gray!10},
	    major grid style={line width=.2pt,draw=gray!50},
	    xlabel={Iteration},
	    ylabel={Cross Entropy},
	  ]
	    \addplot[blue!25] table [x=iteration, y=loss, col sep=comma] {figs/learning_curve_xp_cable.csv};
	    \addplot[red] table [x=iteration, y=smooth_loss, col sep=comma] {figs/learning_curve_xp_cable.csv};

	  \addlegendentry{Raw}
	  \addlegendentry{Smoothed}

		\end{axis}

		\end{tikzpicture}
	    \caption{Evolution of the training loss over the coaxial cable during the first 500 iterations}
	    \label{fig:lc_cable}
\end{figure}

The evolution of the \gls{CE} as a function of the number of training iterations is shown in Fig.~\ref{fig:lc_cable} for the first 500 iterations of training over a coaxial cable.
The \gls{CE} serves as loss function for training.
It can be seen that after only a few hundred of iterations, the loss has decreased by three orders of magnitude.
With our setup based on consumer class hardware, training for a few hundred iterations takes only a few minutes.

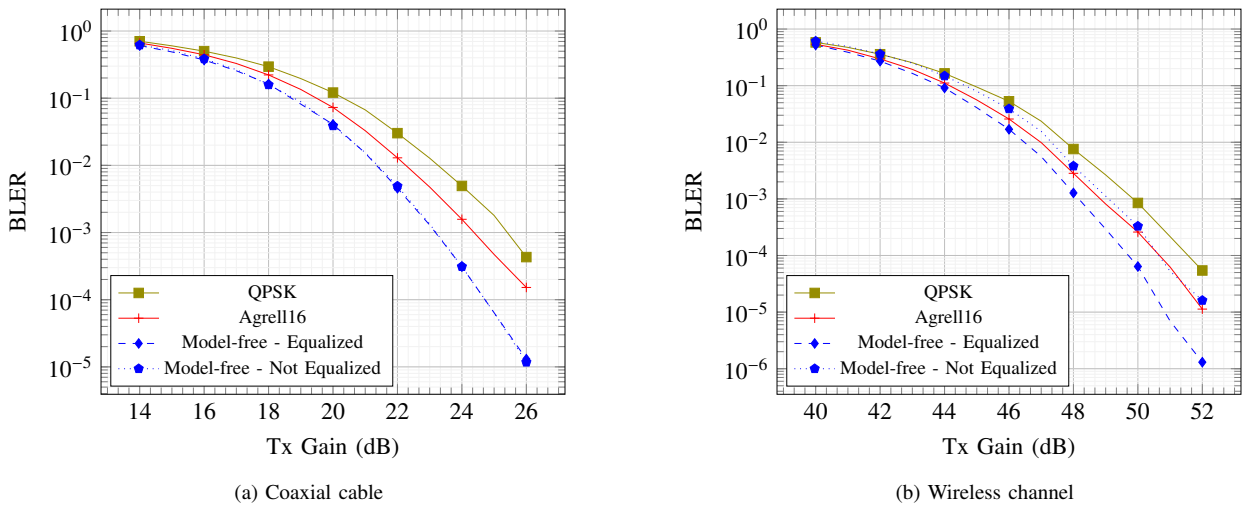
\begin{figure*}
    \centering
    \begin{subfigure}{0.45\linewidth}
	\begin{tikzpicture}[scale=0.9]
	  \begin{axis}[
	    ymode=log,
	    grid=both,
	    grid style={line width=.4pt, draw=gray!10},
	    mark options={solid,scale=1},
	    major grid style={line width=.2pt,draw=gray!50},
	    minor tick num=5,
	    xlabel={Tx Gain (dB)},
	    ylabel={BLER},
	    legend style={at={(0.02, 0.02), font=\footnotesize},anchor=south west},
	    mark repeat={2}
	  ]
	    \addplot[olive, mark=square*] table [x=tx_gain, y=qpsk, col sep=comma] {figs/xp_cable.csv};
	    \addplot[red, mark=+] table [x=tx_gain, y=agrell256, col sep=comma] {figs/xp_cable.csv};
	    \addplot[dashed, blue, mark=diamond*] table [x=tx_gain, y=alt_eq, col sep=comma] {figs/xp_cable.csv};
	    \addplot[dotted, blue, mark=pentagon*] table [x=tx_gain, y=alt_neq, col sep=comma] {figs/xp_cable.csv};

	  \addlegendentry{QPSK}
	  \addlegendentry{Agrell16}
	  \addlegendentry{Model-free - Equalized}
	  \addlegendentry{Model-free - Not Equalized}

		\end{axis}

		\end{tikzpicture}
	    \caption{Coaxial cable}
	    \label{fig:xp_cable}
	\end{subfigure} \qquad
    \begin{subfigure}{0.45\linewidth}
	\begin{tikzpicture}[scale=0.9]
	  \begin{axis}[
	    ymode=log,
	    grid=both,
	    grid style={line width=.4pt, draw=gray!10},
	    mark options={solid,scale=1},
	    major grid style={line width=.2pt,draw=gray!50},
	    minor tick num=5,
	    xlabel={Tx Gain (dB)},
	    ylabel={BLER},
	    legend style={at={(0.02, 0.02), font=\footnotesize},anchor=south west},
	    mark repeat={2}
	  ]
	    \addplot[olive, mark=square*] table [x=tx_gain, y=qpsk, col sep=comma] {figs/xp_wc.csv};
	    \addplot[red, mark=+] table [x=tx_gain, y=agrell256, col sep=comma] {figs/xp_wc.csv};
	    \addplot[dashed, blue, mark=diamond*] table [x=tx_gain, y=alt_eq, col sep=comma] {figs/xp_wc.csv};
	    \addplot[dotted, blue, mark=pentagon*] table [x=tx_gain, y=alt_neq, col sep=comma] {figs/xp_wc.csv};

	  \addlegendentry{QPSK}
	  \addlegendentry{Agrell16}
	  \addlegendentry{Model-free - Equalized}
	  \addlegendentry{Model-free - Not Equalized}

		\end{axis}

		\end{tikzpicture}
	    \caption{Wireless channel}
	    \label{fig:xp_wc}
	\end{subfigure}

	\caption{\gls{BLER} of different schemes over a coaxial cable and wireless channel}
	\label{fig:xp_res}
\end{figure*}

The \gls{BLER} of various schemes over a coaxial cable and wireless channel are compared in Fig.~\ref{fig:xp_res}.
The alternating algorithm outperforms \gls{QPSK} in both settings.
For the coaxial cable (Fig.~\ref{fig:xp_cable}), it also outperforms Agrell.
Manual equalization leads to identical performance as with the transformer network.
On the wireless channel (Fig.~\ref{fig:xp_wc}), the autoencoder with prior equalization achieves better performance than Agrell, while it does not without.
The reason for this is under investigation, but we expect that it could be resolved with longer training time, or different \gls{NN} architectures.
Compared to the results presented in~\cite{dorner2017deep}, where no gains over a well-designed baseline were obtained experimentally, we demonstrate here clear benefits of training over the actual channel with the proposed alternating training algorithm.

\section{Discussion and Outlook}

We have introduced an alternating algorithm for training of autoencoder-based communication systems without a channel model.
Our simulation results show that an autoencoder trained with the proposed method achieves the same performance as when trained using usual backpropagation assuming perfect knowledge of the channel model.
The universality of the alternating algorithm was illustrated by showing that it is able to train complex autoencoders, e.g., for joint source-channel coding, and enables the same performance as model-aware training on a simplified fiber-optical channel model.
We have also presented experimental results of the first prototype of an autoencoder-based communication system trained directly over actual channels.
In contrast to prior work, our setup achieves gains over well-known baselines.
The main reason for this is that transmitter and receiver can be jointly optimized for the actual channel.

A drawback of the alternating algorithm is the requirement of a feedback link at training, which might be unpractical in some settings.
It was also observed that training of the transmitter using the proposed gradient approximation method does not scale with the number of channel uses.
This is due to the variance of the loss function gradient estimator, which increases with the number of channel uses.
However, we believe that this drawback can be addressed using variance reduction techniques, such as~\cite{greensmith2004variance}.
Another open issue of training of communication systems over actual channels is the relatively long coherence time compared to the rate at which examples can be processed at training.
As a consequence, only a few channel realizations are observed (at best) over a single minibatch, leading to poor sample efficiency and slow (if any) convergence.
This is an important obstacle to training communication systems that are able to generalize well to a wide range of channel realizations, which may be solved by artificially introducing channel variations, e.g., by training on channel emulators.

 \appendices

\rev{
\section{Proof of Theorem~\ref{tm:lim_sigma}\label{app:proof}}
}

We denote by $\Xc$ and $\Yc$ the domains of $\hat{\pi}_{\bar{\xv},\sigma}$ and $p(\yv|\xv)$ respectively.
The proof of Theorem~\ref{tm:lim_sigma} is given in \rev{Appendix~\ref{sec:notcompact}.}
In \rev{Appendix~\ref{sec:compact}}, the case where $\Xc$ and $\Yc$ are compact sets is considered.
Under this assumption, which is valid for any hardware implementation, the sufficient conditions for the result to hold take a much simpler form as for the general case.

\subsection{$\Xc$ and $\Yc$ are not compact}\label{sec:notcompact}

Assume the following conditions hold true:
\begin{itemize}
	\item[(C1)] $\hat{\pi}_{\bar{\xv}, \sigma}$ is invariant to translation, i.e., if $\xv \thicksim \hat{\pi}_{\bar{\xv},\sigma}$ then $\xv + \varepsilonv \thicksim \hat{\pi}_{\bar{\xv}+\varepsilonv,\sigma}$.

	\item[(C2)] $\hat{\pi}_{\bar{\xv}, \sigma}$ converges to the Dirac distribution as $\sigma$ decreases towards $0$, i.e, given a continuous and bounded function $g$: $\lim_{\sigma \to 0} \EE_{\xv \thicksim \hat{\pi}_{\bar{\xv},\sigma}} \LP g(\xv,\yv) \RP = g(\bar{\xv},\yv)$.

	\item[(C3)] There exists a function $h : (\xv, \yv) \mapsto \RR^{2N}$ integrable \gls{wrt} $\yv$ and such that
\begin{equation}
	\forall \bar{\xv},~\forall \yv,~\big| \EE_{\xv \thicksim \hat{\pi}_{\bar{\xv},\sigma}} \LP \nabla_{\zv}p(\yv|\zv)|_{\zv = \xv} \RP \big| \leq h(\bar{\xv},\yv)	
\end{equation}
where the inequality holds element-wise, and $\zv$ was introduced for readability.
\end{itemize}
Note that (C1) and (C2) are valid for a normal distribution with mean $\bar{\xv}$ and standard deviation $\sigma$.
Intuitively, (C3) requires that the gradient of the channel distribution \gls{wrt} $\xv$ smoothed by averaging over $\xv$, using $\hat{\pi}_{\bar{\xv},\sigma}$ as weights, is dominated by some function independent of $\sigma$ and integrable \gls{wrt} $\yv$.

One can check by direct calculation that
\begin{align} \label{eq:grad_eq}
	&\EE_{\xv} \LP \nabla_{\bar{\xv}} \log{\pi_{\bar{\xv},\sigma}(\xv)} p(\yv|\xv) \RP \nonumber \\
	&= \nabla_{\bar{\xv}} \EE_{\xv} \LP p(\yv|\xv) \RP\\
	&= \EE_{\xv} \LP \nabla_{\zv} p(\yv|\zv)\lvert_{\zv = \xv} \RP
\end{align}
where $\xv \thicksim \hat{\pi}_{\bar{\xv},\sigma}$.
The first equality is based on the log-trick, while the second equality is obtained through the change of variable $\varepsilonv = \xv - \bar{\xv}$ and using the translation invariance of $\hat{\pi}_{\bar{\xv},\sigma}$ (C1).
Note that the exchange of integration and differentiation requires regularity conditions, discussed, for example, in~\cite{l1995note}.
From the convergence of $\hat{\pi}_{\bar{\xv},\sigma}$ to the Dirac distribution (C2), it follows that
\begin{equation}
	\lim_{\sigma \to 0} \EE_{\xv} \LP \nabla_{\zv} p(\yv|\zv)\lvert_{\zv = \xv} \RP
	=  \nabla_{\zv} p(\yv|\zv)\big|_{\zv = \bar{\xv}}.
\end{equation}
Intuitively, this equality means that, by relaxing the transmitter output to a random variable, the channel gradient can be approximated with arbitrarily good precision.
One can combine the gradient of $\widehat{\Lc}$ given in (\ref{eq:L_hat_grad}) and (\ref{eq:grad_eq}) as
\begin{multline}
	\nabla_{\thetav_T}\widehat{\Lc}
	= \EE_m \Bigg\{ \nabla_{\thetav_T} f^{(T)}_{\thetav_T}(m) \int l \LB f^{(R)}_{\thetav_R}(\yv), m \RB\\
  \cdot \EE_{\xv \thicksim \hat{\pi}_{f^{(T)}_{\thetav_T}(m),\sigma}} \LP \nabla_{\zv} p(\yv|\zv)\lvert_{\zv = \xv} \RP d\yv \Bigg\}.
\end{multline}
Since $l$ is bounded and (C3) holds true, we can conclude by Lebesgue's dominated convergence theorem that
\begin{multline}
	\lim_{\sigma \to 0} \nabla_{\thetav_T}\widehat{\Lc}
		= \EE_m \Bigg\{ \nabla_{\thetav_T} f^{(T)}_{\thetav_T}(m) \int l \LB f^{(R)}_{\thetav_R}(\yv), m \RB\\
    \cdot \nabla_{\zv} p(\yv|\zv)\big|_{\zv = f^{(T)}_{\thetav_T}(m)} d\yv \Bigg\} = \nabla_{\thetav_T}\Lc
\end{multline}
where the last equality is due to (\ref{eq:tx_grad}).
This concludes the proof.

\subsection{$\Xc$ and $\Yc$ are compact} \label{sec:compact}

Assume that $\Xc$ and $\Yc$ are compact sets and (C1) and (C2) hold true.
We require the following additional conditions:
\begin{itemize}
	\item[(C4)] $\nabla_{\xv} p(\yv|\xv)$ is continuous jointly in $\xv$ and $\yv$.
	\item[(C5)] $\hat{\pi}_{\bar{\xv}, \sigma}$ \emph{uniformly} converges to the Dirac distribution as $\sigma$ decreases towards $0$, i.e,\\
  $\forall \eta > 0,~\exists \sigma_0 > 0,~\forall \sigma \leq \sigma_0,~\forall \bar{\xv},~\forall \yv:$:
  \begin{equation} \label{eq:lm_h}
  \abs{\int_{\Xc} \hat{\pi}_{\bar{\xv},\sigma}(\xv) g(\xv,\yv) d\xv - g(\bar{\xv},\yv)} \leq \eta.
  \end{equation}
\end{itemize}
The required conditions on $\hat{\pi}_{\bar{\xv}, \sigma}$ are similar to the ones required in~\cite[Appendix]{silver2014deterministic}.
As it was noticed by the authors of this work, any function $h : \RR^n \mapsto \RR$ that is continuously differentiable, supported on a compact set $\Cc$, and with unit total integral can be used to construct a distribution $\hat{\pi}_{\bar{\xv},\sigma}(\xv)$ that satisfies (C1) and (C5): $\hat{\pi}_{\bar{\xv},\sigma}(\xv) = \frac{1}{\sigma^n} h(\frac{\xv - \bar{\xv}}{\sigma})$.

\begin{lmm}
Assuming (C1), (C2), (C4), (C5) hold true, then

\begin{equation}
    \lim_{\sigma \to 0} \nabla_{\thetav_T} \widehat{\Lc}(\thetav_T, \thetav_R) = \nabla_{\thetav_T} \Lc(\thetav_T, \thetav_R).
\end{equation}

\end{lmm}
\begin{IEEEproof}
Because $\Xc$ and $\Yc$ are compact sets, and $\nabla_{\xv} p(\yv|\xv)$ is assumed to be continuous (C4), from the boundedness theorem, there exists a strict finite bound of $\abs{\nabla_{\xv} p(\yv|\xv)}$ denoted by $M$.
Define
\begin{equation}
	\eta \triangleq \min_{\xv \in \Xc, \yv \in \Yc} \LP M - \Big| \nabla_{\zv}p(\yv|\zv)\lvert_{\zv=\xv} \Big| \RP.
\end{equation}
As $M$ is strict, it follows that $\eta > 0$.
Because $\hat{\pi}_{\bar{\xv},\sigma}(\xv)$ uniformly converges to the Dirac distribution as $\sigma$ decreases (C5), there exists $\sigma_0$ such that $\forall \sigma \leq \sigma_0,~\forall \bar{\xv} \in \Xc$,
\begin{equation}
	\Big| \EE_{\xv} \LP \nabla_{\zv} p(\yv|\zv)\lvert_{\zv=\xv} \RP
	- \nabla_{\zv} p(\yv|\zv)\big|_{\zv=\bar{\xv}} \Big| \leq \frac{\eta}{2}
\end{equation}
leading to, for any $\sigma \leq \sigma_0$, $\bar{\xv} \in \Xc$, and $\yv \in \Yc$,
\begin{align}
	&\Big| \EE_{\xv} \LP \nabla_{\zv} p(\yv|\zv)\lvert_{\zv=\xv} \RP \Big| - \Big| \nabla_{\zv} p(\yv|\zv)\big|_{\zv = \bar{\xv}} \Big| \nonumber\\
  \leq &\Big| \EE_{\xv} \LP \nabla_{\zv} p(\yv|\zv)\big|_{\zv = \xv} \RP - \nabla_{\zv} p(\yv|\zv)\big|_{\zv = \bar{\xv}} \Big| \leq \frac{\eta}{2}
\end{align}
and, therefore,
\begin{equation}
	\Big| \EE_{\xv} \LP \nabla_{\zv} p(\yv|\zv)\lvert_{\zv=\xv} \RP \Big| < M.
\end{equation}
Hence, $\Big| \EE_{\xv} \LP \nabla_{\zv} p(\yv|\zv)\lvert_{\zv=\xv} \RP \Big|$ is dominated by $M$ for $\sigma \leq \sigma_0$.
Moreover, because $\Yc$ is compact, $\int_{\Yc} M d\yv < \infty$.
Therefore, the conditions required in \rev{Appendix~\ref{sec:notcompact}} hold true, which concludes the proof.
\end{IEEEproof}

\rev{
\section{Gradient estimation with SPSA}\label{app:spsa}
}

\rev{
A different line of work~\cite{Raj2018} proposed the use of \gls{SPSA}~\cite{spall1992multivariate} to address the missing gradient problem.
With this approach, gradient descent is performed using the following gradient substitute at each iteration $k$
\begin{equation} \label{eq:spsa}
  \gv \triangleq \frac{\Lc(\thetav_T + c_k \mathbf{\Delta}, \thetav_R) - \Lc(\thetav_T - c_k \mathbf{\Delta}, \thetav_R)}{2c_k} \mathbf{\Delta}
\end{equation}
where $\{c_k\}$ is a decreasing sequence of positive real numbers, and $\mathbf{\Delta}$ is a random vector following a Rademacher distribution.
Despite the promising results  in~\cite{Raj2018}, where transmitters of low complexity were considered, we experimentally found that this approach fails to scale to a large number of trainable parameters $\thetav_T$.
This is because (\ref{eq:spsa}) does not take advantage of the knowledge of $f^{(T)}_{\thetav_T}$ when estimating the loss function gradient, as opposed to our method which only needs to estimate the gradient of $p(\yv|\xv)$ \gls{wrt} to its inputs $\xv$.
}
\rev{
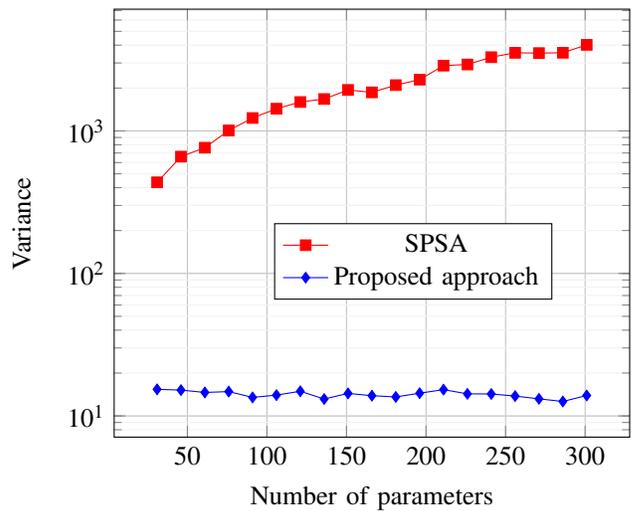
\begin{figure}
  \centering
  \begin{tikzpicture}
    \begin{axis}[
      ymode=log,
      grid=both,
      grid style={line width=.4pt, draw=gray!10},
      major grid style={line width=.2pt,draw=gray!50},
      xlabel={Number of parameters},
      ylabel={Variance},
      legend style={at={(0.85,0.5)}},
    ]
      \addplot[red, mark=square*] table [x expr=\thisrowno{0}*3+1, y=spsa, col sep=comma] {figs/var_spsa_rl.csv};
      \addplot[blue, mark=diamond*] table [x expr=\thisrowno{0}*3+1, y=rl, col sep=comma] {figs/var_spsa_rl.csv};
    \addlegendentry{SPSA}
    \addlegendentry{Proposed approach}
    \end{axis}
    \end{tikzpicture}
      \caption{Evolution of the gradient estimator variance with the number of units}
      \label{fig:spsa_rl_var}
\end{figure}
To illustrate this effect, let us consider the simple regression problem
\begin{equation}
  \underset{\psiv}{\text{minimize}}~J(\psiv) \triangleq \mathbb{E}_{a} \LP \LB f_{\psiv}(a) - a^2 \RB^2 \RP
\end{equation}
where $a$ follows a uniform distribution on $[0,1]$, and the function $f_{\psiv}$ should try to approximate the mapping $a \mapsto a^2$.
\rev{Assume that $f_{\psiv}$ is a \gls{NN} made of two dense layers which use biases, the first with $m$ units and ReLu activation, the second with a single unit and linear activation.}
We choose $\hat{\pi}_{\bar{\xv}, \sigma}$ to be a normal distribution with mean $\bar{\xv}$ and variance $\sigma^2$.
For both \gls{SPSA} and the proposed approach, the variance of the estimators used to perform gradient descent are numerically evaluated for a batch size of $S = 1000$ samples, and over 1000 initializations of the \gls{NN} parameters $\psiv$,
\rev{with $\sigma$ set to $0.1$ and the \gls{SPSA} hyper-parameters values suggested in~\cite{Raj2018}.}
No training was performed so that the variance was estimated after the random \gls{NN} initialization.
This numerical evaluation is performed for a number of $m$ units in the first layer varying from 5 to 100, so that the total number of parameters is $3m+1$.
Fig.~\ref{fig:spsa_rl_var} shows the impact of the number of parameters on the estimators' variances.
The variance is significantly larger using \gls{SPSA} and, moreover, increases linearly with the number of parameters.
This is not the case of the proposed approach, which achieves much lower variance which also does not increase with the number of parameters.
As a consequence of high gradient estimator variance, \gls{SPSA} struggles to train complex transmitter architectures with a large number of parameters.
}


\bibliographystyle{IEEEtran}
\bibliography{IEEEabrv,bibliography}

\begin{IEEEbiography}
    [{\includegraphics[width=1in,height=1.25in,clip,keepaspectratio]{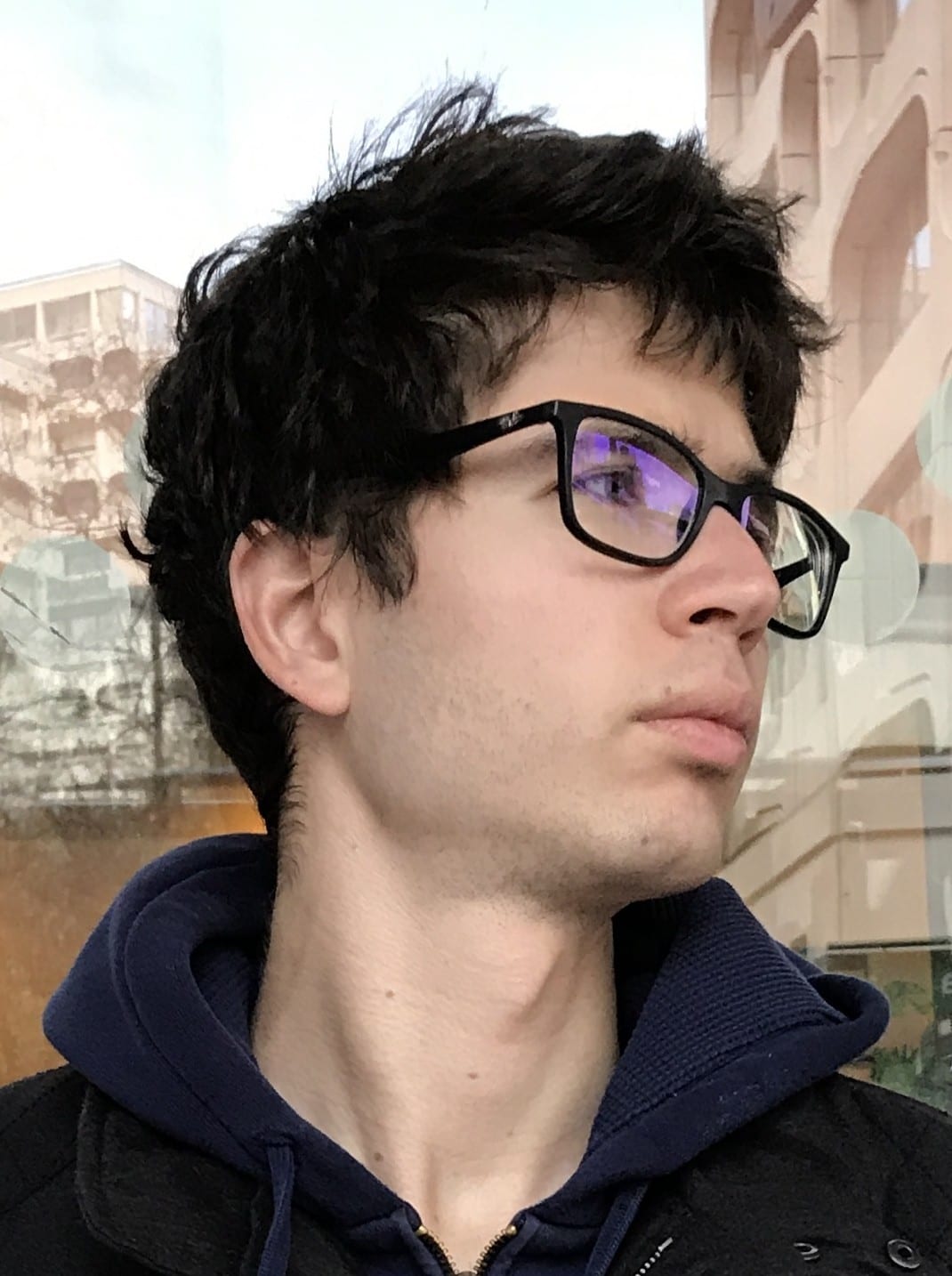}}]{Fayçal Ait Aoudia}
received the M.Sc degree in computer science engineering from the National Institute of Applied Sciences of Lyon (INSA Lyon) in 2014, and the Ph.D degree in signal processing from the University of Rennes 1 in 2017. His Ph.D work focused on energy management and communication protocols design for autonomous wireless sensor networks.  Since 2017, he is a Researcher at Nokia Bell Labs France, where he works on machine learning for wireless communications. He has received the 2018 Nokia AI Innovation Award.
\end{IEEEbiography}

\begin{IEEEbiography}
    [{\includegraphics[width=1in,height=1.25in,clip,keepaspectratio]{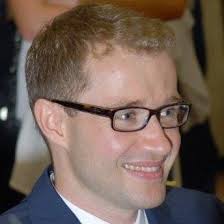}}]{Jakob Hoydis} (S'08 -- M'12 -- SM'19)
received the diploma degree (Dipl.-Ing.) in electrical engineering and information technology from RWTH Aachen University, Germany, and the Ph.D. degree from Supelec, Gif-sur-Yvette, France, in 2008 and 2012, respectively. He is a member of technical staff at Nokia Bell Labs, France, where he is investigating applications of deep learning for the physical layer. Previous to this position he was co-founder and CTO of the social network SPRAED and worked for Alcatel-Lucent Bell Labs in Stuttgart, Germany. His research interests are in the areas of machine learning, cloud computing, SDR, large random matrix theory, information theory, signal processing, and their applications to wireless communications. He is a co-author of the textbook "Massive MIMO Networks: Spectral, Energy, and Hardware Efficiency" (2017). He is recipient of the 2018 Marconi Prize Paper Award, the 2015 Leonard G. Abraham Prize, the IEEE WCNC 2014 best paper award, the 2013 VDE ITG Forderpreis, and the 2012 Publication Prize of the Supelec Foundation. He has received the 2018 Nokia AI Innovation Award and has been nominated as an Exemplary Reviewer 2012 for the IEEE Communication Letters. He is currently chair of the IEEE COMSOC Emerging Technology Initiative on Machine Learning for Communications.
\end{IEEEbiography}

\pagebreak
\end{document}